# Phonon-Mediated Chirality Transfer from Organic Cation to Inorganic lattice in Hybrid Perovskites

Sankaran Ramesh[1,2*], Prasenjit Mandal[1], Rafael Araujo[3], Jakob Thyr[3], Pratik Bhagwat[1], Yong Li[1], Tomas Edvinsson[3], Tönu Pullerits[1*], Dmitry Baranov[1*]

[1]Division of Chemical Physics and NanoLund, Department of Chemistry, Lund University, P.O. Box 124, Lund, SE-221 00 Sweden

[2]Centre for Interdisciplinary and Convergent Technologies, Indian Institute of Technology Kharagpur, West Bengal, India-721302

[3]Department of Materials Science and Engineering, Solid State Physics, Uppsala University, P.O. Box 35, Uppsala SE-751 03, Sweden

*Corresponding authors: sankar@atdc.iitkgp.ac.in, tonu.pullerits@chem.lu.se, dmitry.baranov@chem.lu.se

## Abstract:

Two-dimensional hybrid metal halide perovskites combine spin-orbit coupling, soft lattice dynamics, and molecular tunability, making them promising platforms for chiral optoelectronics. While chiral organic spacers are known to induce optical activity in the inorganic framework, the microscopic mechanism by which molecular chirality couples to the inorganic lattice is actively studied. Here, we investigate coherent vibrational dynamics in chiral ($R$-MBA)$_2$PbI$_4$ and its racemic analogue using circular-polarized transient absorption spectroscopy. A vibrational mode at ~5.7 *meV* is present in the chiral material and not observed in the racemic counterpart, confirmed as a lattice phonon by agreement between time-domain and steady-state measurements. Ab-initio calculations reveal correlated libration of the organic spacer and bending displacement of the Pb-I framework analogous to a classical Wilberforce pendulum whose sense is locked to the structural handedness of the lattice. Polarization-resolved measurements reveal a transient circular dichroism whose exact sign reversal between ($R$-MBA)$_2$PbI$_4$ and ($S$-MBA)$_2$PbI$_4$ is consistent with a periodic modulation of the excitonic rotational strength driven by the coupled lattice displacement. These results establish dynamic chirality transfer through the coupled librational-bending mechanism, with direct implications for phonon-driven chiral optical responses and spin-selective transport.

## Introduction

Chiral metal–halide perovskites (MHPs) sit at the intersection of molecular chirality, strong spin–orbit coupling, and soft lattice dynamics, making them compelling candidates for next-generation optoelectronic and spintronic devices. Among these materials, two-dimensional (2D) Ruddlesden-Popper type perovskites are of particular interest owing to their excellent optical properties and exceptional compositional tunability.[1,2] They feature a natural quantum-well-like layered structure, with the band edges located on the inorganic metal halide layer, and strong dielectric confinement is afforded by a tunable chiral organic spacer. These unique characteristics govern their electronic properties, such as an exciton binding energy in the 200-300 meV range.[3,4]

In chiral 2D MHPs, the chiral organic spacer can induce optical activity at the excitonic transition, indicating that molecular chirality influences the inorganic metal-halide framework.[5-8] Structural descriptors based on bond vectors have been introduced to break down chirality into components of organic spacers, in and out-of-plane metal halide framework, and hydrogen bonding connecting the organic and inorganic parts.[9] Such description provided insights into temperature-dependence of structural chirality, finding that structural chirality of organic spacers is more robust to thermal disorder than that of inorganic framework. Specifically, with increasing temperature the chirality of inorganic framework between *R/S*-$MBA_2PbI_4$ enantiomers quickly diminishes. These structural descriptors reflect time-averaged configurations. All reported pathways for chirality transfer describe static or quasi-static structural mechanisms.[10] It is important to obtain experimental insight into the dynamical processes by which molecular chirality is imprinted onto the inorganic framework.

Lattice vibrations provide a natural dynamical channel through which chirality encoded in the organic spacer can be transferred to the inorganic framework and its associated electronic states. Recent theoretical work[11,12] has proposed the existence of chiral phonons - lattice vibrations carrying angular momentum and breaking improper rotational symmetry[13,14] - in chiral MHPs, suggesting a possible dynamical pathway linking molecular chirality to electronic and spin degrees of freedom.[15-18] An alternative, not mutually exclusive possibility is chirality transfer through lattice vibrations where librations of the spacer and translational displacement of inorganic framework are coupled through the crystal potential, analogous to a classical Wilberforce pendulum,[19,20] in which linear and torsional oscillations exchange energy through internal geometry of the oscillator.

Distinguishing these mechanisms experimentally is challenging, particularly at low phonon energies where the relevant modes are densely spaced and short-lived.[21,22] Coherent phonon spectroscopy[23,24] provides a route to resolve exciton-phonon coupling directly in the time domain: it has revealed strong exciton-phonon coupling and excitonic fine structure in 2D MHPs[25,26] but direct connections between coherent lattice dynamics and chirality in these systems have not yet been established.

In this work, we investigate ultrafast vibrational dynamics of the organic and inorganic sublattices in the chiral perovskites (*R/S*-MBA)$_2$PbI$_4$ and its racemic analogue using transient absorption spectroscopy, complemented by Raman spectroscopy and first-principles calculations. By directly comparing chiral and achiral systems, we experimentally identify vibrational signatures sensitive to circular polarization, discuss their association with structural chirality, and clarify the role of lattice dynamics in mediating chirality transfer and chiral optical responses in hybrid perovskites.

## Results

### Chiral Perovskite Films

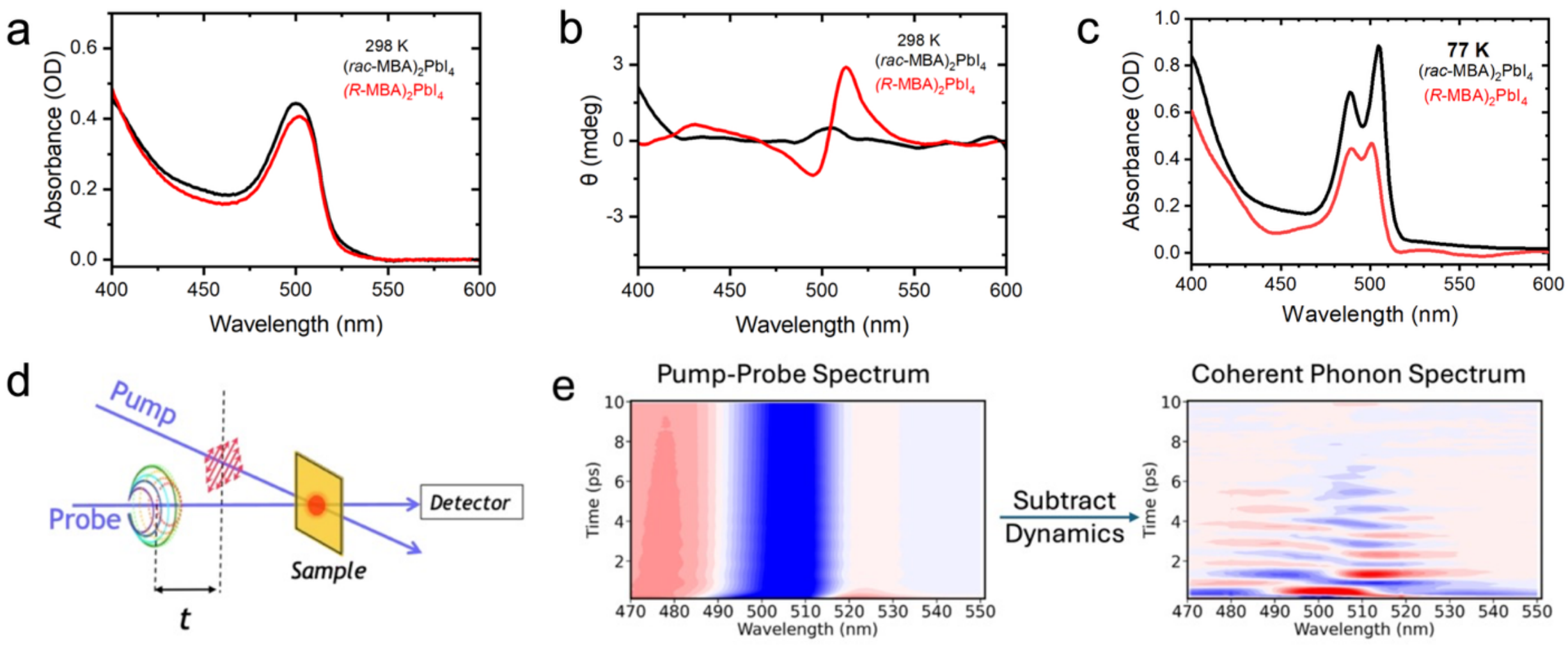

Figure 1: Optical characterization of (*R*-MBA)$_2$PbI$_4$ and (*rac*-MBA)$_2$PbI$_4$ thin films and Transient Absorption (TA) analysis of coherent phonons. (a) UV–Vis absorption spectra showed an excitonic peak at ~502 nm for both films. (b) CD spectra revealed a strong signal at ~515 nm for the chiral (*R*-MBA)$_2$PbI$_4$ film, while no CD response was observed for the racemic sample. (c) Absorption spectra of films at 77 K; splitting of the excitonic peak is observed. (d) Schematic of the TA experiment using linearly-polarized pump pulse with a wavelength of 400 nm and a circularly-polarized broadband (white light) probe pulse. (e) Illustration of the workflow of TA data analysis showing steps of dynamics subtraction to obtain spectra of coherent phonons.

Thin films of (*R*-MBA)$_2$PbI$_4$ and (*rac*-MBA)$_2$PbI$_4$ were deposited from solution by spin-coating[27] and have similar morphology (**Figures S1**). The details of the sample preparation, along with a structural and steady-state optical characterization, are provided in the Methods section. These chiral 2D MHPs have alternating chiral organic and achiral inorganic structural units. The X-ray diffraction patterns of the two films exhibit periodic peaks (**Figure S2**) indicating the periodicity of the structure along the stack of organic-inorganic layers. **Figure 1a** shows the absorption spectra of the ($R$-MBA)$_2$PbI$_4$ and ($rac$-MBA)$_2$PbI$_4$ films. A sharp peak at ~502 nm is the optical signature of an excitonic transition. **Figure 1b** shows the circular dichroism (CD) spectra of the two samples. The chiral perovskite ($R$-MBA)$_2$PbI$_4$ exhibits a double peak feature near the absorption edge, a signature of optical activity also known as the Cotton effect.[28] The ($rac$-MBA)$_2$PbI$_4$ shows no CD, as expected from a non-chiral material. The absorption spectra of the films at 77 K reveal splitting of the excitonic resonance into two peaks, as shown in **Figure 1c** further discussed in the Results section.

## Transient Absorption Spectroscopy

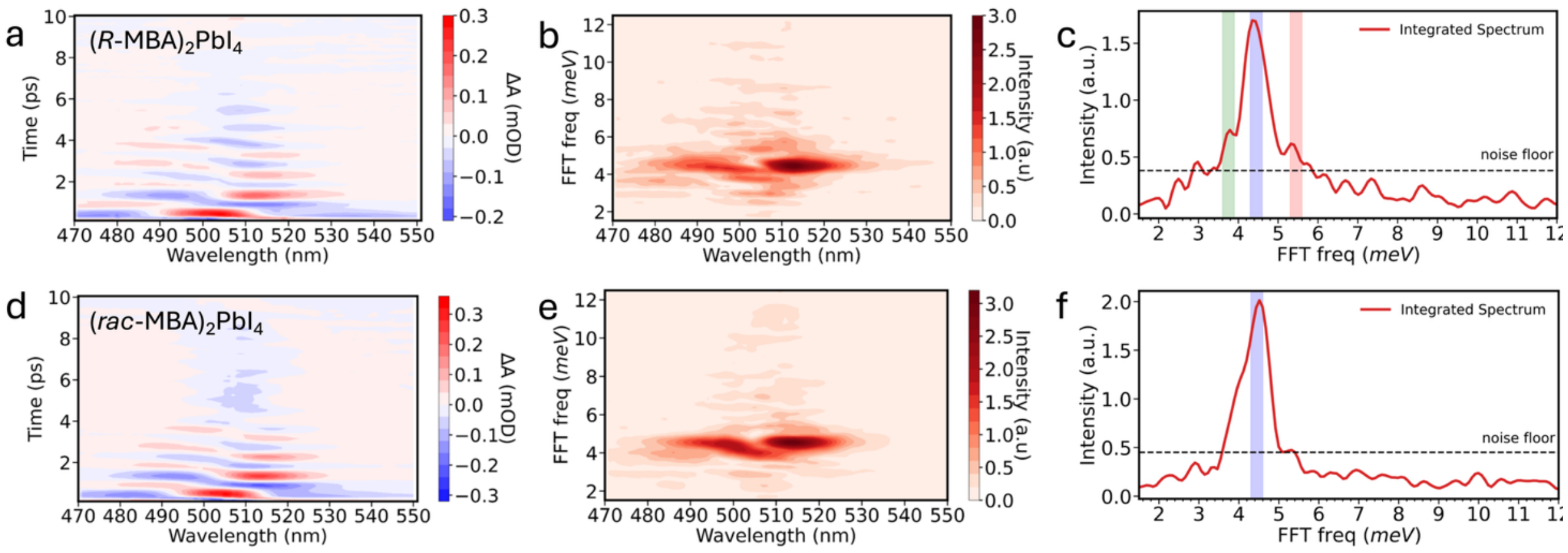

Figure 2: Room temperature comparison of coherent phonons in ($R$-MBA)$_2$PbI$_4$ (a-c) and ($rac$-MBA)$_2$PbI$_4$ (d-f) measured with linear-polarized (LP) pump and left circularly-polarized (LCP) probe. (a) Residual oscillatory signal in the transient absorption spectrum of ($R$-MBA)$_2$PbI$_4$ obtained after subtracting the non-oscillatory decay. (b) Fourier Transform of the temporal map in (a). (c) Integrated signal intensity along probe wavelength of the frequency map (b). (d-f) same as (a-c) for ($rac$-MBA)$_2$PbI$_4$.

To study the time-resolved vibrational dynamics in the two systems, we employed transient absorption (TA) spectroscopy with linear-polarized (LP) pump pulses and circular-polarized (CP) probe pulses (**Figure 1d**) at room temperature (298 K). The films were excited by linearly polarized 400 nm, 80 fs laser pulses with a fluence of 8 μJ/cm$^2$. A time-delayed supercontinuum pulse employed as the probe measures the differential absorption signal. Further details of the experimental setup are provided in the Methods section.

The pump pulse creates excitations in the continuum of the electronic band. In 2D MHPs, the excitations are known to form to strongly bound excitonic states.[4,29] Due to strong exciton-phonon coupling, excitation by the pump pulse also generates coherent vibrational wavepackets that modulate the differential absorption. [26,30] Some of us recently investigated coherent phonons in lead-free MHPs and revealed distinct vibrational signatures associated with charge–lattice coupling.[31,32] The origin of the spectral modulations is explained by the deformation potential of the ionic lattice, creating a periodic modulation of the electronic bandgap of the system.[31,32]

Here, the periodic modulations of the differential absorption caused by the LP pump and detected by a CP probe were extracted and analyzed (**Figure 1e**). The left panel shows a typical TA spectrum collected in the experiment. The spectrogram shows negative signal band that is assigned the ground state bleaching (GSB). It also has two positive photo-induced absorption bands (PIA) on either side of the GSB. These signals decay over several ps and reflect the relaxation of the exciton to the ground state. To study the vibrational signals due to exciton-phonon coupling, these non-oscillatory signals in the TA spectrum were removed by fitting the dynamics with multi-exponential functions and extracting the residual oscillations (right panel). **Figure S3** shows the collected TA spectra of the films at 298 K and **Figure S4** illustrates examples of this procedure implemented for chosen probe wavelengths.

**Figure 2a** shows the residual spectrum extracted from the TA spectrum of a (*R*-MBA)$_2$PbI$_4$ film. Pronounced oscillations are observed close to the excitonic resonance, which coincides with the ground state bleach (GSB) signal in the TA spectrum (**Figure S3**). The oscillation pattern is characterized by an exponential amplitude decay in time and an abrupt change in phase of oscillations at the GSB band along the wavelength axis. These are typical features of a coherent phonon-induced modulation of probe absorption.[31] The energy spectra of the oscillations are obtained by Fourier transforming this signal, and are shown in **Figure 2b**. It has been reported that optical phonons of the Pb-I octahedra have frequencies between 2-6 meV.[33] We therefore focused on the region > 2 meV here, and signals at lower energies were suppressed by applying a high-pass frequency filter. Distinct features are observed in this region of the energy spectra, which we therefore assign to coherent optical phonon modes.[11] **Figure 2c** shows the integrated Fourier-transformed signal across the wavelength range. The dashed horizontal line indicates the threshold intensity for confidently detecting peaks which have been defined as three times above statistical noise (Details in **Supporting Information Section 4**). The results of TA experiment on a racemic sample (*rac*-MBA)$_2$PbI$_4$ are shown **Figure 2d-f** for comparison. The phonon modes are described below ordered by their energy:

Table 1: Summary of coherent phonon modes observed in the chiral and racemic 2D perovskite at 298 K, as shown in Figure 2c and Figure 2f

| Phonon Mode | Peak position | |
|---|---|---|
| | (*R*-MBA)$_2$PbI$_4$ | (*rac*-MBA)$_2$PbI$_4$ |
| M1 | 3.7 ± 0.2 meV | - |
| M2 | 4.4 ± 0.2 meV | 4.4 ± 0.2 meV |
| M3 | 5.4 ± 0.2 meV | - |

The M2 peak indicates a phonon mode that is observed in both samples at 4.4 ± 0.2 meV. While a peak at M1 is distinctly observed in ($R$-MBA)$_2$PbI$_4$, it may also be present in in ($rac$-MBA)$_2$PbI$_4$, where it is observed as a shoulder to the M2 peak. M3 is observed only in the ($R$-MBA)$_2$PbI$_4$ sample and not observed in ($rac$-MBA)$_2$PbI$_4$ (**Figure 2e-f**). This reflects differences in exciton-phonon coupling between the two samples. To clarify this, **Figure S5** shows the integrated FFT spectra of the chiral and racemic samples overlaid on each other. Additional experiments were carried out with linear and right-circular probe polarizations for comparison of observed frequencies and are shown in **Figures S6** and **S7**, without noticeable shifts within the 0.3 meV resolution limit of our Fourier transform (Details of FFT resolution limit are given in **Supporting Information Section 5**).

To clarify the presence and nature of the phonon modes observed at room temperature, the TA experiments were repeated at 77 K. Both samples now show two GSB peaks at ~490 nm and ~502 nm (**Figure S8**). Considering the ~60 meV energetic separation of the two peaks, their origin cannot be due to phonon replica, since phonon modes of the inorganic unit typically have energy < 60 meV. Further, the two GSBs have different dynamics, indicating that they originate from two distinct transitions. Here we label the two states as exciton A and B, respectively in order of their energy.

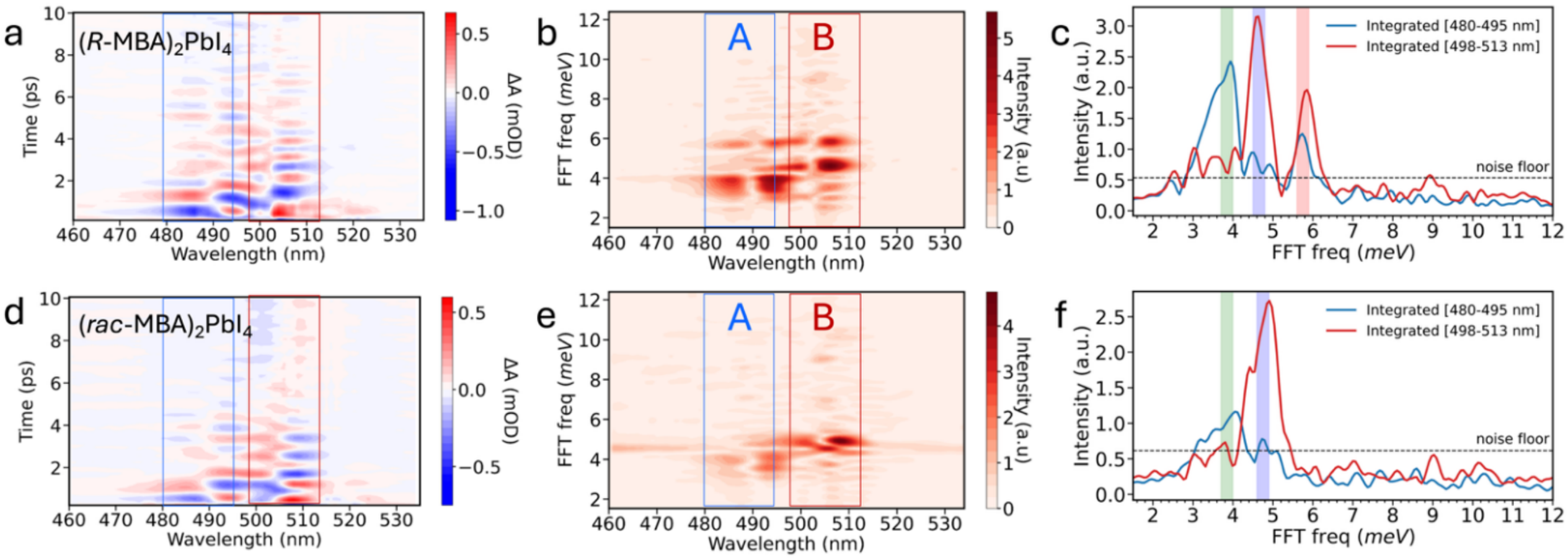

Figure 3: Low temperature (77 K) Comparison of coherent phonons in ($R$-MBA)$_2$PbI$_4$ (a-c) and ($rac$-MBA)$_2$PbI$_4$ (d-f). Measured with linear-polarized (LP) pump and left circularly-polarized (LCP) probe. (a) Residual oscillatory signal in the transient absorption spectrum of ($R$-MBA)$_2$PbI$_4$ obtained after subtracting the non-oscillatory decay. (b) Fourier Transform of the temporal map in (a). Blue and red rectangles indicate the spectral region of exciton A and exciton B, respectively. (c) Integrated signal intensity along probe wavelength of the frequency map (b). (d-f) same as (a-c) for ($rac$-MBA)$_2$PbI$_4$.

The oscillatory signals in the 77 K TA data for $(R\text{-MBA})_2PbI_4$ and $(rac\text{-MBA})_2PbI_4$ are shown in **Figure 3a,d**, extracted in the same way as for room temperature data (**Figure S8**). **Figure 3b,e** shows the energy spectra of oscillations, and **Figure 3c,f** shows the spectra integrated over the resonance wavelengths of excitons A and B, along with the detection threshold level. The phonon modes prominently observed are described in **Table 2,** ordered by their energy:

Table 2: Summary of coherent phonon modes observed in the chiral and racemic 2D perovskite at 77 K, as shown in Figure 3c and Figure 3f. Bracketed terms indicate which exciton (A, B) couples to each phonon.

| Phonon Mode | Peak position | |
|---|---|---|
| | $(R\text{-MBA})_2PbI_4$ | $(rac\text{-MBA})_2PbI_4$ |
| M1 | 3.8 ± 0.2 meV (A) | 3.8 ± 0.2 meV (A) |
| M2 | 4.6 ± 0.2 meV (B) | 4.6 ± 0.2 meV (B) |
| M3 | 5.7 ± 0.2 meV (A,B) | - |

The modes (M1, M2, M3) identified from the frequency maps are confirmed from the temporal analysis of the coherent phonon signals, given in **Figure S12, S13.** The signal in the time domain can be reproduced by a sum of decaying oscillations of modes, and the analysis also clarifies the dephasing timescales and the relevant mechanism by which the coherent phonons are generated. The time-domain analysis also identifies additional weak components whose FFT amplitude falls below the detection threshold (a ~3.0 meV component in $(R\text{-MBA})_2PbI_4$ at 77 K and a ~3.7 meV component in $(rac\text{-MBA})_2PbI_4$ at 298 K**)**. Comparing the two methods, while the integrated FFT is conservative and the temporal analysis is more sensitive. The latter however requires assuming the number of oscillating components. We adopt here mainly the conservative picture provided by the integrated FFT. The time-domain analysis is provided in **Supporting Information Section 6.**

Lowering the temperature reveals further details of excitonic structure and exciton-phonon coupling. First, the emergence of double peak is not a result of structural phase transition, as no structural changes are reported to occur for these compounds at these temperatures.[34] Second, the phonon modes M1, M2 and M3 are also detected at 77 K, but appear to be blue shifted (higher energy) compared to the observation at room temperature. The observed blueshift with decreasing temperature is consistent with a lattice stiffening upon thermal contraction.[34-36] Third, different phonon modes couple to excitons A and B. In $(R\text{-MBA})_2PbI_4$, it is observed that exciton A couples to modes at M1 and M3, while exciton B couples to M2 and M3 modes. In $(rac\text{-MBA})_2PbI_4$, Exciton A exhibits weak coupling to lattice vibrations, with contributions mainly from M1 mode. In contrast, Exciton B couples mainly to a mode at M2. The states A and B are ascribed to excitons with distinct polaronic character. The observations of this study are consistent with reports of co-existence of distinct excitonic states in 2D MHPs with different vibrational coupling.[25,26]

The stronger coupling of exciton B to the lattice in (*rac*-MBA)$_2$PbI$_4$, relative to the more comparable A/B coupling in (*R*-MBA)$_2$PbI$_4$, could reflect either (i) a difference in exciton-phonon coupling between the two structures, or (ii) a difference in the orbital character of excitons A and B between the two structures, without requiring the phonon landscape itself to differ. Considering the different crystal structures of both samples, which will be discussed further in the following section, the observed difference in A/B coupling strength is attributable, at least in part, to a different phonon landscape (A detailed discussion on the coupling of A and B states to the lattice is given in **Supporting Information Section 7**).

Changing the probe polarization does not noticeably change observed vibration frequencies, similar to room temperature measurements (**Figure S10, S11**). This observation is consistent with recent Raman optical activity measurements on chiral amino acid crystals, where the splitting was either below instrument resolution or obscured by overlapping modes, even as the modes exhibited opposite intensities for LCP and RCP excitation.[37] It is worth noting that the polarization of the pump pulse in all our experiments is linear, hence there is no net angular momentum transferred from the photon to the material. This may be limiting the observed contrast. A stronger probe-helicity dependence of the coherent phonon-induced signal may be expected under such a circularly polarized pump.[38]

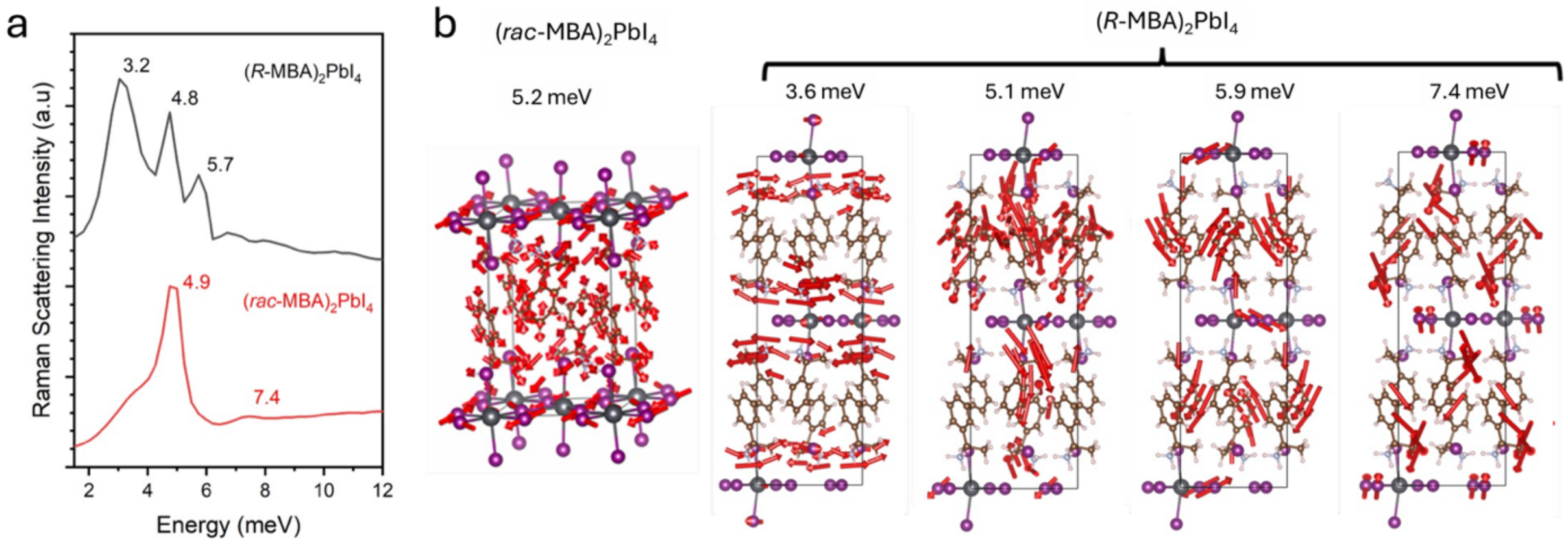

Figure 4: (a) Low temperature (78-80 K) Raman spectra of (*R*-MBA)$_2$PbI$_4$ and (*rac*-MBA)$_2$PbI$_4$ thin films. Peaks at 3.2 meV, 4.8 meV and 5.7 meV are consistent with the modes M1 (3.8 ± 0.2 meV), M2 (4.6 ± 0.2 meV), M3 (5.7 ± 0.2 meV) identified in from the TA experiments for (*R*-MBA)$_2$PbI$_4$ (**Table 2**). (b) Representative Raman-active vibrational modes associated with the experimentally observed low-energy features in the samples.

In summary, the chiral ($R$-MBA)$_2$PbI$_4$ sample couples to mode M3 that has no clearly observed analogue in the ($rac$-MBA)$_2$PbI$_4$ sample, consistent with the room temperature observation. This is further cross-verified with steady-state Raman spectroscopy at 78 K, which yielded results consistent with the TA. The films were excited with a Raman laser at 532 nm and a filter with a cut-off at 10 cm$^{-1}$ was used, enabling the clear resolution of modes as low as 3 meV. The results are shown in **Figure 4a**. In ($R$-MBA)$_2$PbI$_4$ three Raman peaks are observed, consistent with the M1, M2 and M3 peaks in the integrated FFT spectra (**Table 2**). The Raman spectra of ($rac$-MBA)$_2$PbI$_4$ is also consistent with the TA results, showing a strong peak near the M2 energy with a shoulder covering the M1 peak region. The results further identify the M3 mode as observed only in the chiral sample. In the next sections we investigate how the mode could be linked to chirality propagation across the organic-inorganic interface.

## Coupled Motion of Organic and Inorganic Layer

The experiments in previous section revealed the M3 phonon mode that couples to excitons which show optical activity in ($R$-MBA)$_2$PbI$_4$. This motivates a theoretical analysis to understand what makes the M3 mode distinct from the rest of the modes. ($R$-MBA)$_2$PbI$_4$ and ($rac$-MBA)$_2$PbI$_4$ form different crystal structures (orthorhombic, $P2_12_12_1$ space group and monoclinic, $P2_1/c$ space group respectively). Ab-initio calculations of the phonon dispersion and theoretical investigation of Raman susceptibilities of modes were performed (**Supporting Information Section 8**) to explain the contrasting exciton-phonon coupling landscape of the two films and to clarify the nature of the modes beyond the simple emergence of additional vibrational modes in ($R$-MBA)$_2$PbI$_4$ as a result of distinct crystal structures. The phonon spectra along and associated individual contributions of atomic species (**Figure S14**) shows that both samples have a dense phonon spectrum. For the studied energy range the Pb and I atoms have a more dominant contribution to the phonon modes and a minor contribution from the atoms on the organic spacer.

Calculations of Raman tensors identified a single strong Raman-active mode at 5.24 meV in ($rac$-MBA)$_2$PbI$_4$, which we assign to the M2 mode observed experimentally in the racemic material (**Figure S15**). The discrepancy (<0.0006 eV) falls within the expected uncertainty of DFT-based vibrational frequencies where the order of the modes and their intensities can be utilized as additional information for the assignment. This mode links the organic and inorganic units (**Figure 4b**), combining Pb–I–Pb bending within the inorganic layers with motion of the organic cations. In ($R$-MBA)$_2$PbI$_4$, four Raman-active modes (modes shown schematically **Figure 4b**) with computed vibrational energies of 3.6, 5.1, 5.9 and 7.4 meV are identified. From the order of the modes in energy, the 3.6 meV mode is assigned to the M1 mode. The 5.1, 5.9 and 7.4 meV modes display a substantially larger contribution from the organic framework than their racemic counterpart does, with the 5.1 meV mode representing a limiting case in which the Pb–I network contributes minimally (**Figure S16**). This peak is assigned to the M2 mode.

The M3 mode is theoretically assigned as follows: considering the theoretical Raman spectra incorporating realistic broadening at 77 K, the 5.9 meV and 7.4 meV modes contribute additively to a single broadened spectral feature as shown in **Figure S15**. The experimental M3 peak is therefore assigned as the unresolved envelope of these closely spaced modes. Among these, the 5.9 meV mode notably exhibits a librational displacement pattern (**Figure 4b**). The vibrational eigenvector of the mode is dominated by libration in the organic spacer. To illustrate the displacement of the organic and inorganic layers separately for each mode, they are visualized separately with larger unit cell representations and rescaled displacement arrows in **Figure S17-S20.**

Therefore, the crucial distinction among the identified modes is the character of the vibrational eigenvectors associated with the experimental M3 mode. Our first-principles calculations reveal that this mode is dominated by librational displacement patterns originating in the chiral organic spacer and dynamically coupled to bending motion of the Pb–I framework. In contrast, the dominant low-energy modes in both materials primarily involve linear displacements and lack such librational components. Our finding is further reinforced by the discovery of the M3 mode in the other chiral enantiomer, ($S$-MBA)$_2$PbI$_4$, as shown in **Supporting Information Section 9** and **Figure S5**. The absence of the M3 mode in ($rac$-MBA)$_2$PbI$_4$, despite otherwise similar inorganic lattice motifs, indicates that this vibration is not a generic consequence of structural distortion but may be linked to molecular stereochemistry.

## Chirality-dependent Excitonic Transient Circular Dichroism

The identification of a mode with librational-bending character linking the organic and inorganic sublattices in the chiral structure raises the question of whether the coupled displacement modulates the chiroptical character of the transition.

Coherent phonons modulate the energy of the excitonic transition via the deformation potential[31] at the frequency of the phonon along its nuclear co-ordinate $Q$. This leads to modulation of the ground state bleach independent of whether the probe helicity is LCP or RCP. A second mechanism that we propose is that phonons with librational displacement in a chiral structure would periodically modify the rotational strength of the excitonic transition:

$$\mathcal{R} = Im[\mu . m] \quad (1)$$

where $\mu$ and $m$ are the electric and magnetic transition dipole moments respectively. $\mathcal{R}$ governs the differential absorption of LCP and RCP light and is directly proportional to the CD of the transition.[28,39] As the phonon displaces the chiral lattice along $Q$, it periodically modifies the spatial arrangement between the chiral organic spacer and the Pb-I framework, changing the overlap between $\mu$ and $m$. This is plausible in chiral 2D perovskites where lattice strain is expected to modulate chirality.[40,41] The transient CD $\Delta\Delta A$ is defined as the difference between the transient absorption with LCP and RCP probe:

$$\Delta\Delta \boldsymbol{A}(\boldsymbol{t}) = \Delta \boldsymbol{A}_{\boldsymbol{LCP}}(\boldsymbol{t}) - \Delta \boldsymbol{A}_{\boldsymbol{RCP}}(\boldsymbol{t}) \tag{2}$$

As explained in **Supporting Information section 10**,

$$\Delta\Delta \boldsymbol{A}(\boldsymbol{t}) = \frac{\boldsymbol{\partial \mathcal{R}}}{\boldsymbol{\partial Q_n}} \boldsymbol{Q_n}(\boldsymbol{t}).\boldsymbol{\mathcal{L}}(\boldsymbol{\omega}) \tag{3}$$

Where $Q_n$ is the normal-mode coordinate and $\mathcal{L}(\omega)$ is the CD lineshape that is assumed to remain unchanged. Since $\mathcal{R}$ is a pseudoscalar that changes sign under inversion, the signal must therefore undergo an exact sign reversal between (*R*-MBA)$_2$PbI$_4$ and (*S*-MBA)$_2$PbI$_4$. The exact sign reversal between enantiomers should constitute the experimental validation of the rotational strength modulation.

The transient CD signal calculated as defined by Equation (2) for (*R*-MBA)$_2$PbI$_4$ is shown in **Figure 5a**. **Figure 5b** shows the same measurement for (*S*-MBA)$_2$PbI$_4$, the isostructural mirror-image enantiomer exhibiting opposite optical activity compared to (*R*-MBA)$_2$PbI$_4$ (**Figure S21**). Both samples show a short-lived $\Delta\Delta A$ signal near the excitonic resonance with an exact sign reversal between them, as shown clearly in the integrated transient CD spectrum over 490-505 nm, shown in **Figure 5c**. The $\Delta\Delta A$ signal is weak (~ $10^{-4}$ OD) and its enantiomeric sign reversal is consistent with the rotational strength modulation mechanism. **Figure 5d** shows the residual TA signal with LCP probe at the GSB energy, which shows relatively stronger signals, to aid visual comparison to the transient CD signal.

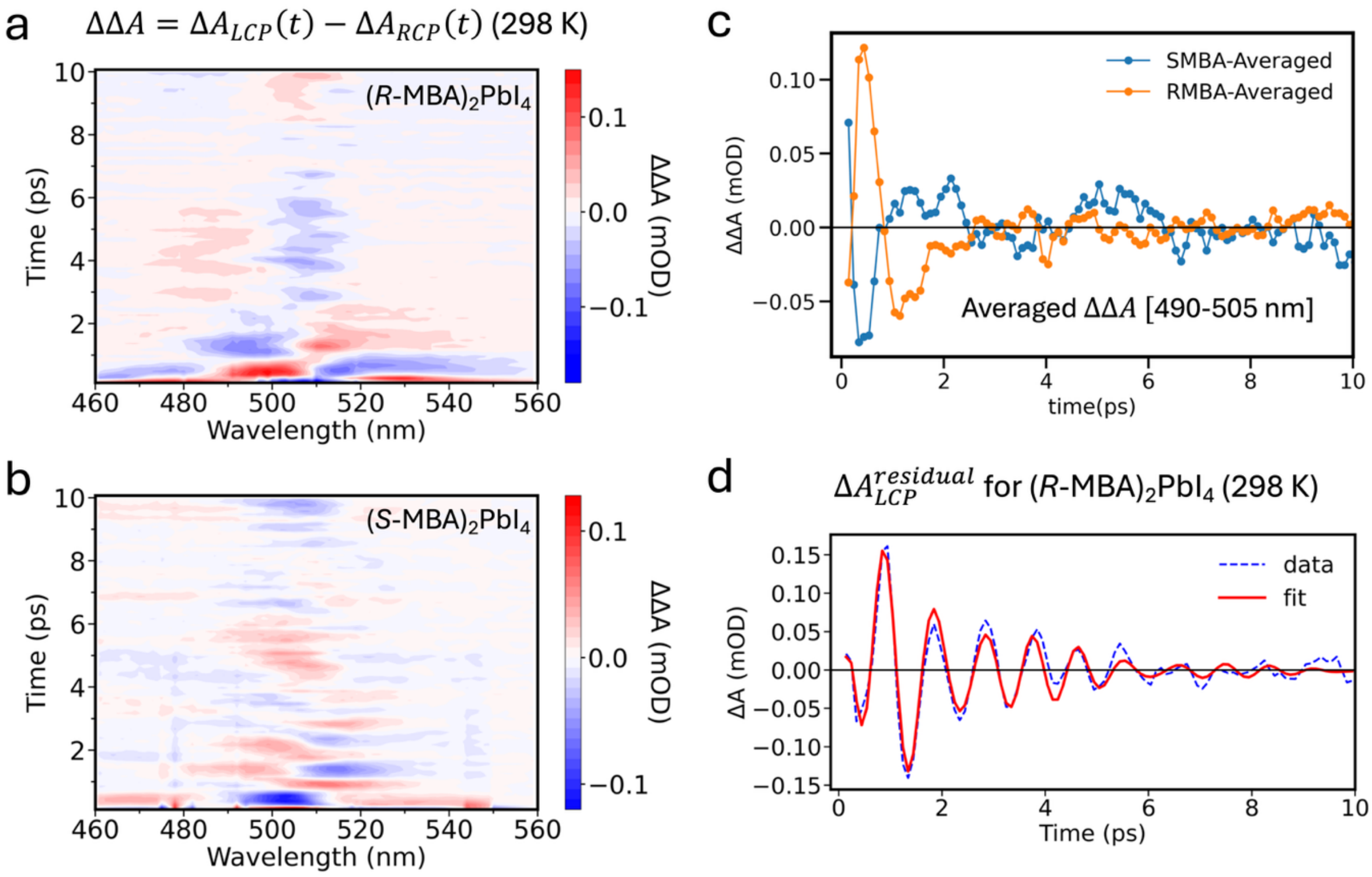

**Figure 5**: Helicity-resolved transient circular dichroism reveals chiroptical phonon coupling in chiral perovskites. (a-b) Pseudo-colour maps of the transient circular dichroism signal $\Delta\Delta A = \Delta A_{LCP}(t) - \Delta A_{RCP}(t)$ as a function of probe wavelength and pump-probe delay at 298 K for (a) $(R\text{-MBA})_2PbI_4$ and (b) $(S\text{-MBA})_2PbI_4$. A short-lived oscillatory signal is observed near the excitonic resonance in both samples, with a dispersive spectral profile consistent with periodic modulation of the excitonic rotational strength. (c) Integrated $\Delta\Delta A$ over 490–505 nm for *R*-MBA (yellow) and *S*-MBA (blue), showing an exact sign reversal between the two enantiomers. The opposite signs demonstrate that the helicity-sensitive coherent oscillation is locked to the structural handedness of the lattice, consistent with the rotational strength modulation mechanism described in the text. (d) Residual transient absorption signal $\Delta A^{residual}$ of $(R\text{-MBA})_2PbI_4$ measured in the LP-LCP configuration at the ground state bleach energy at 298 K, shown for comparison.

## Discussion

Our ab initio calculations identify a mode at ~5.9 meV with librational displacement of the chiral spacer and bending displacement of the Pb-I framework which we assign to the M3 mode (5.7 ± 0.2 meV at 77 K) observed only in the chiral sample. An important question at this point is whether this mode also carries pseudo-angular momentum, making it a chiral phonon. Quantitative analysis of the Raman tensors does not yield a clear signature of non-zero phonon angular momentum at the $\Gamma$-point. Chiral phonons in this class of material would require either a higher-symmetry inorganic framework or finite-q modes, neither of which is accessed by our $\Gamma$-point measurement. We therefore do not claim it to be a chiral phonon in the strict theoretical sense.[13,14]

Instead, we propose that the experimental M3 mode corresponds to a coupled displacement analogous to a classical Wilberforce pendulum, which is a mechanical system in which librational and translational degrees of freedom are coupled through the internal geometry of the oscillator. The libration of the chiral spacer and the bending of the Pb-I bonds, mechanically coupled through the chiral crystal, produces a helical oscillatory displacement whose sense is locked to the structural handedness of the lattice. Our results are consistent with the lattice modes inducing a periodic modulation of rotational strength of the excitonic transition.

Beyond providing a mechanistic picture of sublattice coupling in chiral 2D MHP, these findings carry implications for spin-selective transport in chiral MHPs. The chirality-induced spin selectivity (CISS) effect has been experimentally established in MBA-based perovskites,[42,43] yet its microscopic mechanism remains debated. This work shows that coherent lattice dynamics can serve as a channel for chirality transfer from spacer to excitonic states in the inorganic framework. This complements the established picture of static chirality transfer through hydrogen bonding and octahedral distortions.[10] By periodically modulating the structural chirality and Pb-I bond geometry, these phonons can modulate the potential experienced by charge carriers, enhancing spin-selective scattering beyond what static structure alone predicts.

Future work combining the time-resolved chiroptical approach demonstrated here with spin-sensitive transport measurements or vibrational spectroscopy would provide a direct test of whether the phonon modes identified are intrinsically chiral and how they contribute to spin selectivity. Specifically, circularly polarized pump excitation should impart angular momentum to the system, amplifying transient CD signals and enabling direct measurements of rotational strength modulation coefficient. Raman optical activity measurements applied to these systems can independently test the M3 modes for the sign-reversing intensity difference expected of a chiral phonon.[37] By establishing a concrete link between molecular stereochemistry, lattice motion, and excitonic chiroptical response, this work opens a route toward engineering phonon-driven chiral functionality in hybrid materials.[44-48]

## Experimental Methods

### Chemicals

Lead(II) oxide (PbO, 99.9%), (*R*)-(+)-α-methylbenzylamine (*R*-MBA,115541-25ML, 98%), (*S*)-(−)-α-methylbenzylamine (*S*-MBA,115568-25G, 98%), hydriodic acid (HI, 57% w/w in $H_2O$, 99.9%), hypophosphorous acid solution ($H_3PO_2$, 50 wt. % in $H_2O$), N,N-dimethylformamide (DMF, 99.8%, anhydrous), acetonitrile (ACN, 99.8%, anhydrous), diethyl ether (99%, anhydrous), toluene (99.8%, anhydrous), polystyrene ($M_w$ 288,800; $M_n$ 131,500) were purchased from Sigma-Aldrich. All chemicals were used as received without further purification.

### Synthesis of (*R/S*-MBA)$_2$PbI$_4$ Crystals

The synthesis of (*R/S*-MBA)$_2$PbI$_4$ single crystals was adopted from Makhija et al.[27] In a typical synthesis, PbO (0.558 g, 2.5 mmol) was dissolved in a mixture of 10 mL aqueous HI and 2 mL aqueous $H_3PO_2$ in a 20 mL glass vial by heating at 70 °C under magnetic stirring (400 rpm setting on the stirring plate). For (*R*-MBA)$_2$PbI$_4$ :when the solution became clear yellow, (*R*)-(+)-α-methylbenzylamine (5 mmol, ca. 637 μL) was added using a mechanical 1-1000 μL micropipette. Orange-coloured crystals appeared immediately. For (*S*-MBA)$_2$PbI$_4$ : similar protocol was followed with the (*S*)-(+)-α-methylbenzylamine (5 mmol, ca. 645 μL) added instead. The solution was then heated to 70 °C under continuous stirring at 400 rpm until the crystals fully dissolved, which took approximately 40 minutes, resulting in a clear yellow solution. The solution was allowed to cool to ambient temperature. During cooling, needle-shaped crystals began to appear, and the reaction mixture was left undisturbed overnight (12 hours) to complete the crystallization process. The crystals were then filtered, washed with diethyl ether, dried under vacuum, and stored in a nitrogen-filled glove box for further use.

### Synthesis of (*rac*-MBA)$_2$PbI$_4$ Crystals

The synthesis of (rac-MBA$_2$)PbI$_4$ was adopted from Ahn et al.[5] with modifications. In a typical synthesis, PbO (0.2 g, 0.896 mmol) was dissolved in 6 mL of aqueous HI in a 20 mL glass vial using magnetic stirring. Subsequently, 200 μL (1.57 mmol) of racemic MBA (a 1:1 mixture of the *R*-MBA and *S*-MBA instead of the commercial *rac*-MBA as in the original recipe) was added and immediately caused the formation of a precipitate. Then the vial was heated to 100 °C in an oil bath to redissolve the precipitate. Once the precipitate fully dissolved, the reaction mixture was let to cool to an ambient temperature, resulting in the crystallization of orange needle-like crystals. These crystals were collected by vacuum filtration and rinsed several times with toluene. The final product was dried in a vacuum desiccator for a minimum of two days and subsequently stored in a nitrogen-filled glove box for further use.

### Film Preparation

Thin films of ($R$-MBA)$_2$PbI$_4$ , ($S$-MBA)$_2$PbI$_4$ or ($rac$-MBA)$_2$PbI$_4$ were prepared by dissolving 4 mg of the crystals in 10 μL of DMF and 250 μL of ACN using a vortex mixer. From this solution, 40 μL was used to prepare the films either by drop-casting or spin-coating onto a pre-cleaned 1 cm × 1 cm glass substrate. After film deposition, the samples were annealed at 60 °C for 5 minutes. For spin coating, the solution was dropped onto the glass substrate and spin-coated at 1000 rpm for 10 seconds with an acceleration of 500 rpm/s, followed by 4000 rpm for 30 seconds with an acceleration of 1000 rpm/s. The film preparation was performed under ambient conditions in air. Shortly after the deposition, drop-cast and spin-coated films were encapsulated by spin-coating 40 μL of polystyrene solution (60 mg/mL in toluene) to protect the films from ambient conditions. The spin-coated films encapsulated with polystyrene were used in transient absorption experiments.

**Characterization and spectroscopic studies**

The films were characterized using a range of techniques. X-ray diffraction (XRD) was conducted using a STOE STADI MP instrument with Cu-Kα radiation (1.54 Å) in reflection mode to identify the crystal structure. Room temperature UV-visible absorption spectra of spin-coated samples were recorded using an Agilent 8453 UV-visible spectrometer. The absorption spectra at 77 K were measured in the transient absorption measurement setup described below. The transmitted intensity of the probe beam without and with the sample loaded in an Oxford Instruments cryostat were measured and absorbance was calculated. The films for the XRD, UV-Vis, and transient absorption measurements were prepared by spin-coating. Circular dichroism (CD) spectra were measured with an OLIS DSM 245 CD/CPL Spectrophotometer equipped with a 150 W Xenon lamp. The films for the CD measurements were prepared by drop-casting. The drop-cast films were imaged by optical microscopy by using a Kern optical microscope (model OBN 148) equipped with a digital camera (ODC 861).

**Transient Absorption**

Transient Absorption (TA) measurements were conducted by using a homemade femtosecond pump-probe setup. Laser pulses (8 W, 796 nm, 60 fs, 4 kHz) come out of Solstice (Spectra Physics) amplifier seeded by a femtosecond oscillator (Mai Tai SP, Spectra Physics). The laser output is split into two beams that each pump collinear optical parametric amplifiers (TOPAS-C, Light Conversion). The first one generates 400 nm wavelength pump pulses, while the other generates 1350 nm pulses that are focused onto a $CaF_2$ plate to generate a supercontinuum pulse, which is used as the probe. The pump beam is set to have orthogonal linear polarization with the probe using a Berek compensator in the path of the pump beam. The probe beam is delayed with respect to the pump using a mechanical delay stage, which can provide a maximum possible pump-probe delay of 10 ns. Achromatic quarter-wave plate (Thorlabs AQWP05M-600) was used to set the polarization state of the supercontinuum to left- and right-circular polarization. The excitation density on both chiral and racemic samples is $\sim 6.5 \times 10^{12}\ cm^{-2}$. Then the probe beam is further divided into two parts: the former is focused on the sample overlapping with the pump pulse, and the latter serves as a reference. The sample is loaded in a cryostat from Oxford Instruments for the measurement at 77 K. After passing the sample, the probe beam is collimated again and relayed onto the entrance aperture of a prism spectrograph. The reference beam is directly relayed on the spectrograph. Both the measurement and reference beams are then dispersed onto a double photodiode array, each holding 512 elements (Pascher Instruments). TA spectra were analyzed for delays from 0 ps to 12 ps, with spectra collected every 0.1 ps. Data analysis was performed using custom-written scripts in Python.

**Density Functional Theory Calculations**

First-principles simulations were performed within the framework of density functional theory (DFT) using the Projector Augmented Wave (PAW) approach as implemented in the Vienna *Ab initio* Simulation Package (VASP) .[49,50] Exchange–correlation effects were described using the Perdew–Burke–Ernzerhof (PBE)[51] functional within the generalized gradient approximation (GGA), combined with Grimme's D3 dispersion correction to account for long-range van der Waals interactions.[52,53] A plane-wave kinetic-energy cutoff of 500 eV was used. Brillouin-zone sampling employed a 2 × 2 × 1 mesh for (R-MBA)$_2$PbI$_4$ and a 1 × 2 × 2 mesh for (rac-MBA)$_2$PbI$_4$. Structural relaxations were performed until the residual forces on all atoms were below $10^{-3}$ eV Å$^{-1}$, and electronic self-consistency was converged to $10^{-8}$ eV. Vibrational frequencies and eigenvectors at the Γ point were obtained using density functional perturbation theory (DFPT). Raman intensities were evaluated within the non-resonant Stokes scattering regime using the Placzek approximation, following Porezag et al.[54] The Raman tensor derivatives were computed numerically using finite differences, based on variations of the macroscopic dielectric tensor induced by atomic displacements of ±0.005 Å along each normal-mode coordinate. The resulting Raman activities were broadened using a Gaussian distribution with a full width at half maximum (FWHM) of 7 cm$^{-1}$. Raman intensities were computed at 70 K to account for phonon occupation via Bose–Einstein statistics.

**Raman Spectroscopy**

Raman spectra were acquired with a Renishaw Invia Reflex spectrometer using a frequency-doubled Nd-YAG laser operating at 532 nm, an 1800 lines/mm grating, and a high performance notch filter (Eclipse™) enabling detection of Raman shifts from 10 cm$^{-1}$ from the Rayleigh line. The spectrometer path length (250 mm) and the spectrometer components (slit, grating, and detector pixel density) gave a resolution of 1 cm$^{-1}$. A Linkam™ THMS600 cryo-stage pumped with liquid nitrogen (77 K) was utilized together with a long working distance objective to enable measurements of the samples. The measurements were done at 78 K for (*rac*-MBA)$_2$PbI$_4$ and 80 K for (*R*-MBA)$_2$PbI$_4$. Raw Stokes spectra are reported throughout without division by the Bose–Einstein thermal factor $(n(\omega, T) + 1)$. The correction rescales low-energy intensities but leaves peak positions and the presence or absence of features unchanged, and the mode assignments in this work rest on frequencies rather than absolute intensities.

## Acknowledgements:

The work of PM and DB is supported by the European Innovation Council Pathfinder Challenges project number 101162112 (RADIANT). DB acknowledges the start-up support from the Faculty of Science, Lund University, the Crafoord Foundation (decision 20230942), and the Swedish Research Council (2024-04967). TP acknowledges support by Swedish Foundation of Strategic Research (IS24-0005), Swedish Energy Agency (50709-1), Olle Engkvists Foundation (235-0422) and Swedish Research Council (2021-05207). TE acknowledges support from the Swedish Research Council (2023-05244) and the Swedish Energy Agency (50667-1, P2020-90215).

# Supporting Information

*for*

## Phonon-Mediated Chirality Transfer from Organic Cation to Inorganic lattice in Hybrid Perovskites

Sankaran Ramesh[1,2*], Prasenjit Mandal[1], Rafael Araujo[3], Jakob Thyr[3], Pratik Bhagwat[1], Yong Li[1], Tomas Edvinsson[3], Tõnu Pullerits[1*], Dmitry Baranov[1*]

[1]Division of Chemical Physics and NanoLund, Department of Chemistry, Lund University, P.O. Box 124, Lund, SE-221 00 Sweden

[2]Centre for Interdisciplinary and Convergent Technologies, Indian Institute of Technology Kharagpur, West Bengal, India-721302

[3]Department of Materials Science and Engineering, Solid State Physics, Uppsala University, P.O. Box 35, Uppsala SE-751 03, Sweden

*Corresponding authors: sankar@atdc.iitkgp.ac.in, tonu.pullerits@chem.lu.se, dmitry.baranov@chem.lu.se

## 1. Sample Characterization

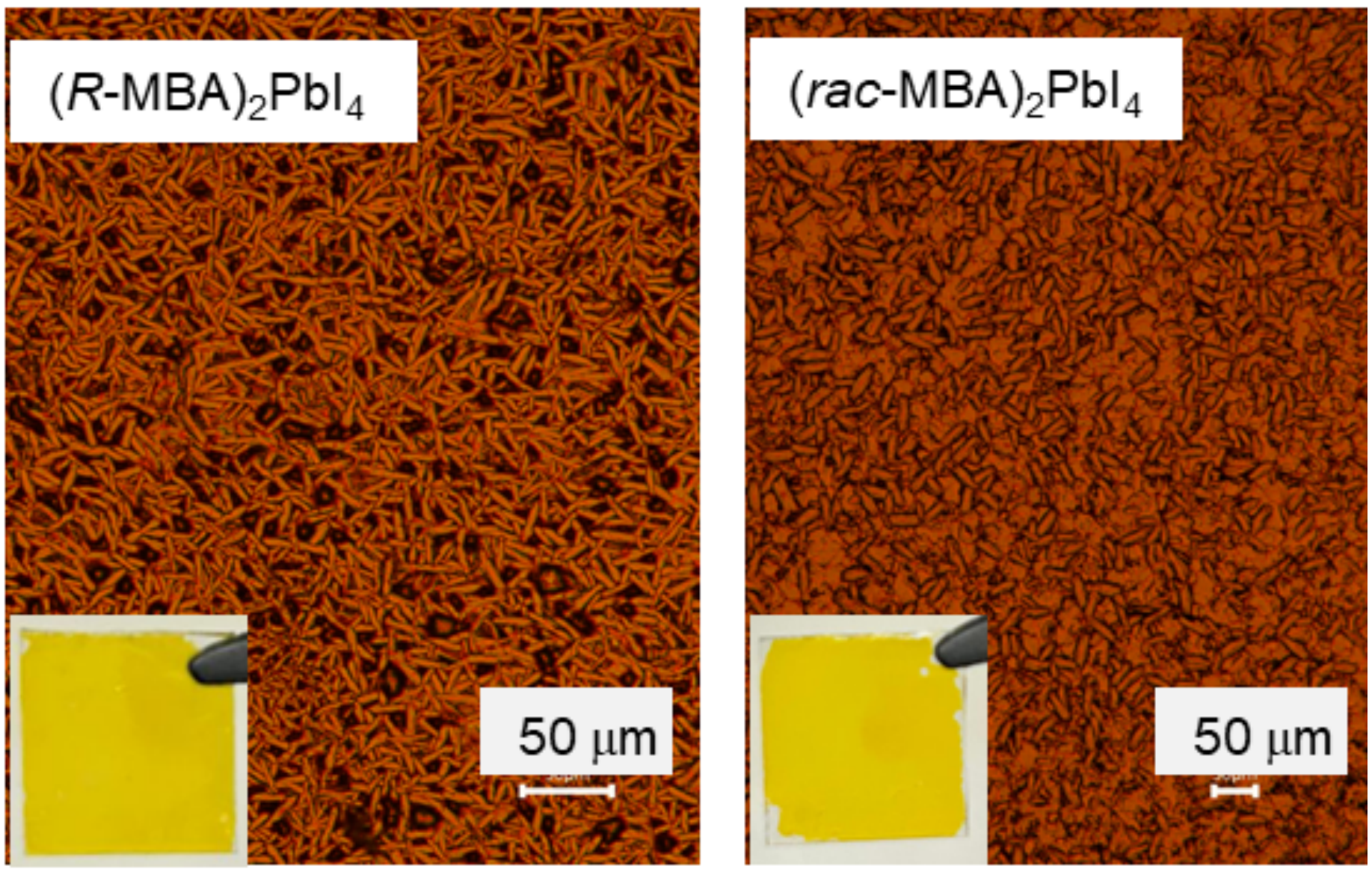

**Figure S1:** Optical microscopy images of the (*R*-MBA)$_2$PbI$_4$ and (*rac*-MBA)$_2$PbI$_4$ thin films, the insets show photographs of the films taken with a smartphone camera under ambient light.

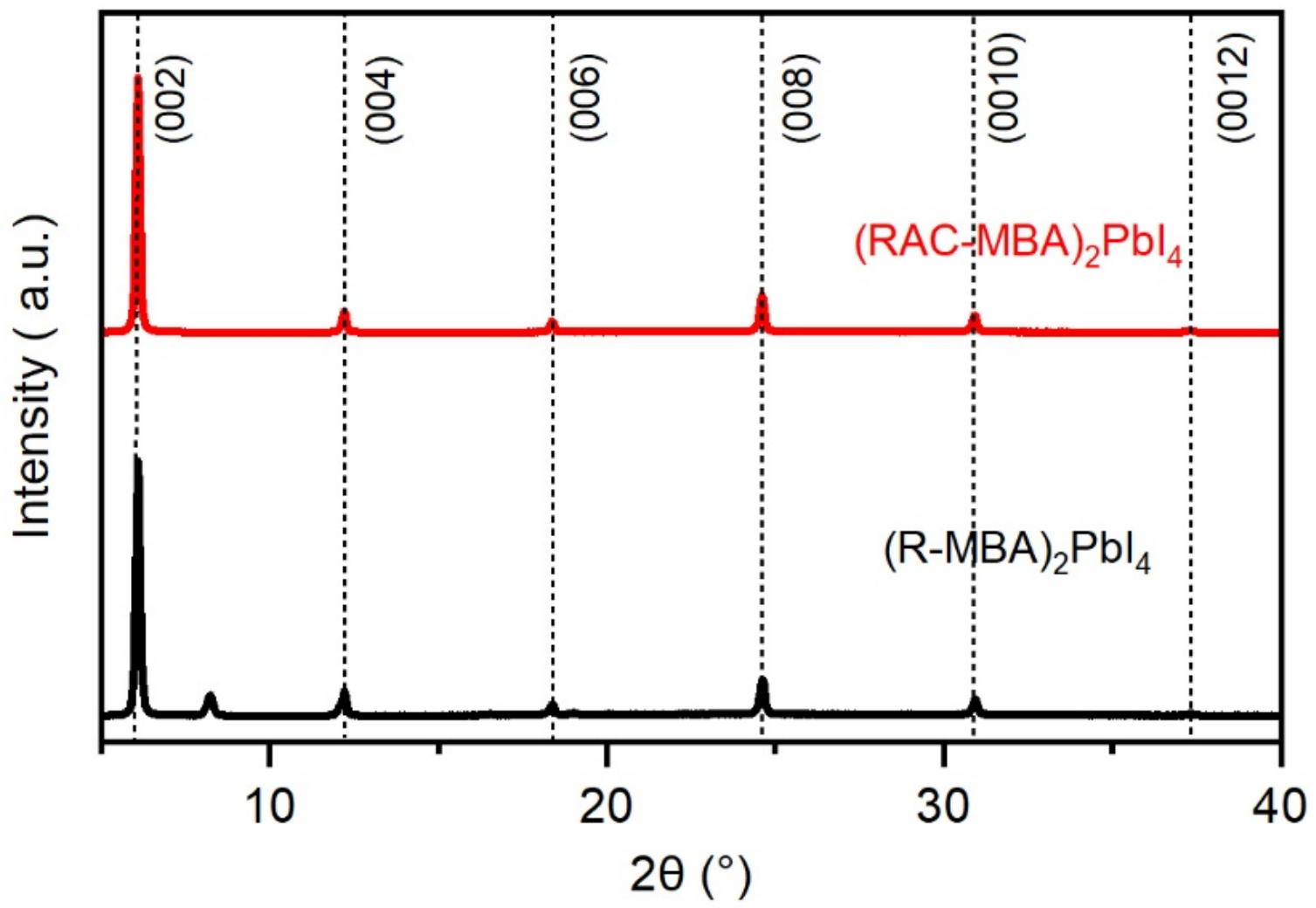

**Figure S2**: XRD patterns of the thin film samples are consistent with those reported in the literature.[1,2] Periodic peaks are observed, indicating their layered structure. The (R-

$MBA)_2PbI_4$ contains an additional peak at 8 degrees that is reported to originate from a minority 1D phase.[1]

## 2. Transient Absorption Experiments at 298 K

Transient Absorption Experiment were conducted at 298 K with 400 nm, 8 μJ/cm$^2$, linear polarized pump pulses and with probe pulses in Left Circular (LCP), Linear (LP) and Right Circular (RCP) polarized configurations to probe the transient circular dichroism signal from the sample. Below are the collected datasets, shown up to 10 ps in different selected probe configurations.

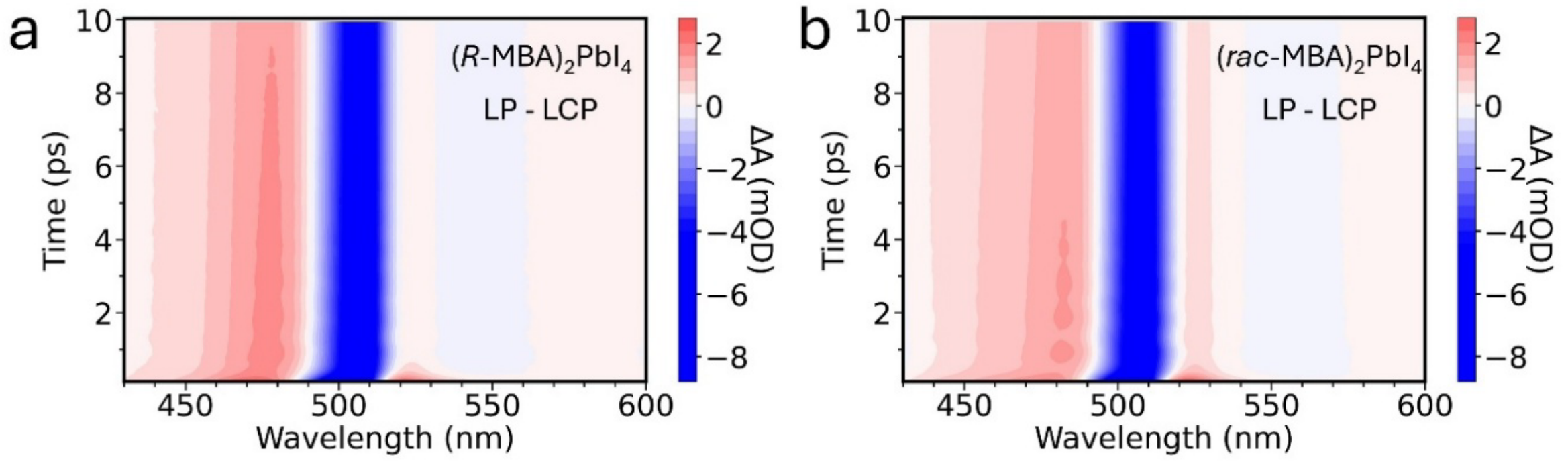

**Figure S3**: Pseudo-colour transient absorption colormap of **(a)** (*R*-MBA)$_2$PbI$_4$ and **(b)** (*rac*-MBA)$_2$PbI$_4$ from 0 to 10 ps using linearly-polarized pump and left circularly-polarized probe pulses.

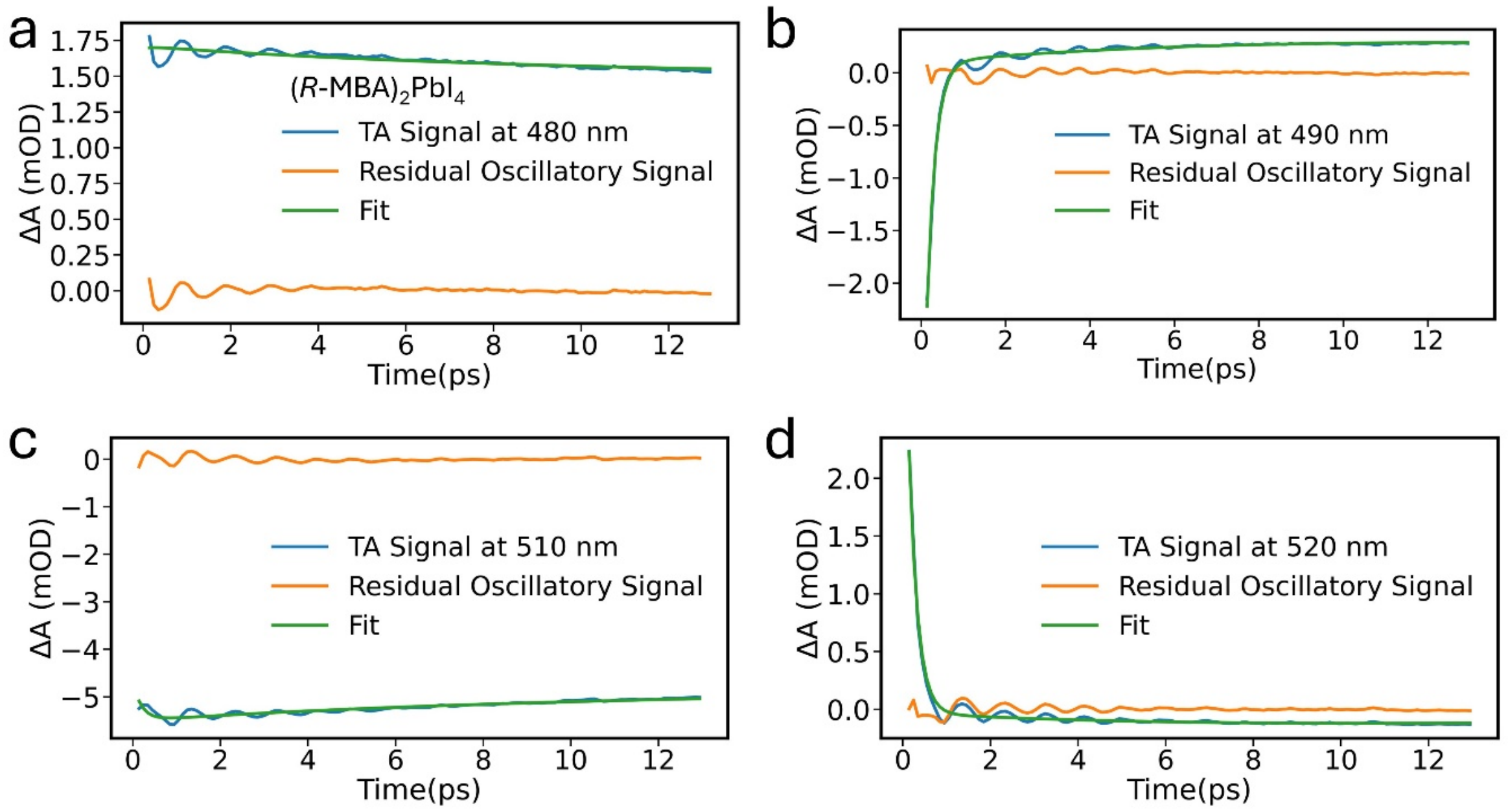

**Figure S4**: Calculation of Residual signal from transient absorption spectra at probe wavelengths for (R-MBA)$_2$PbI$_4$, measured in the LP-LCP configuration at 298 K: (a) 480 nm (b) 490 nm (c) 510 nm and (d) 520 nm

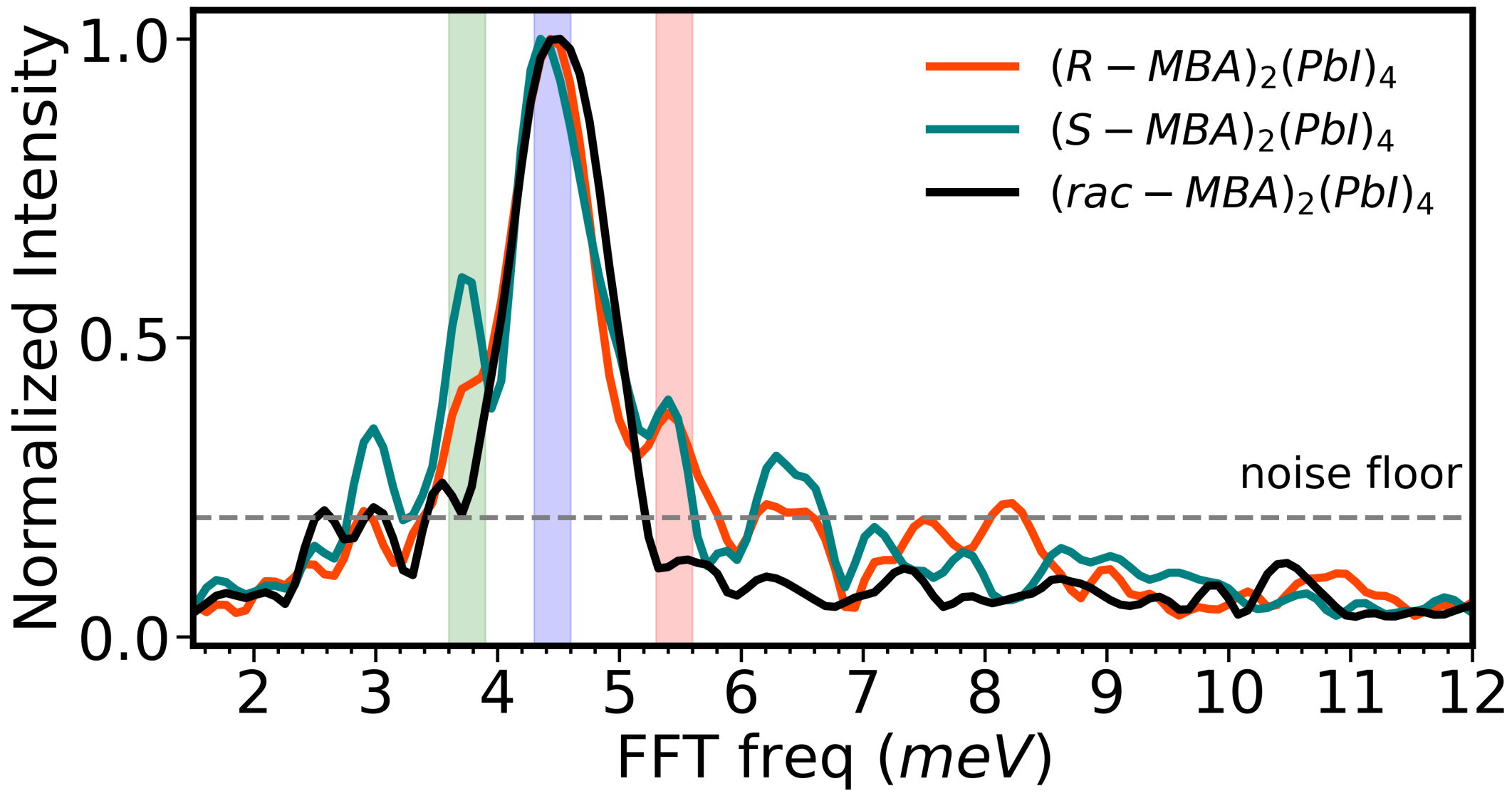

**Figure S5:** Comparison of the FFT of residual TA spectra, normalized, of (*R*-MBA)$_2$PbI$_4$, (*S*-MBA)$_2$PbI$_4$ and (*rac*-MBA)$_2$PbI$_4$. The 5.4 ± 0.2 meV feature is clearly observed above the noise floor in both chiral samples (red shaded region).

## 2.1 Effect of Changing Probe Polarization on Coherent Phonon Signals

In both ($R$-MBA)$_2$PbI$_4$ and ($rac$-MBA)$_2$PbI$_4$ thin films, the vibrational frequencies do not show considerable shift depending on probe pulse polarization. Frequency resolution of the Fourier transformed spectrum is ~0.3 meV.

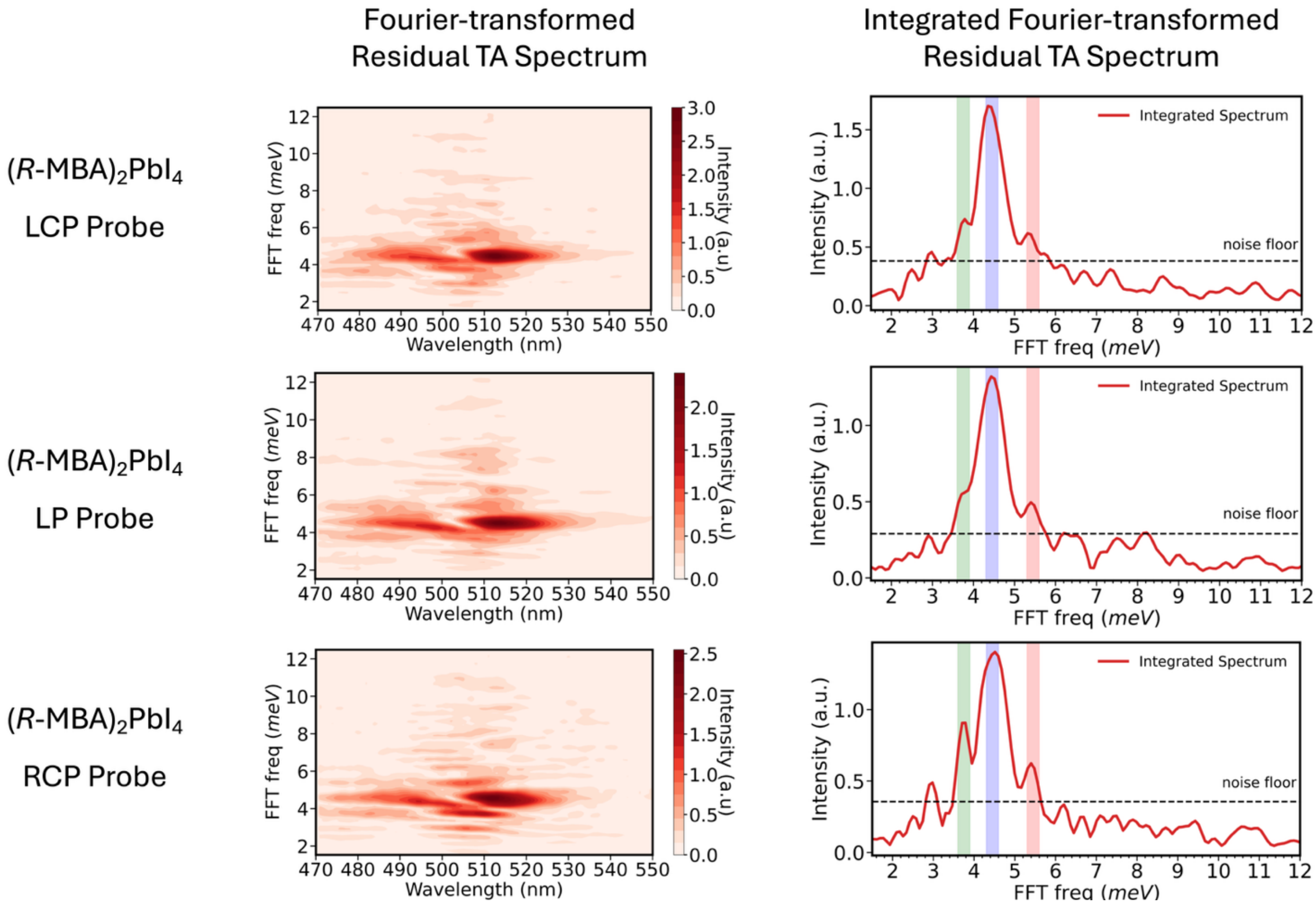

**Figure S6**: Residual oscillatory signal in the transient absorption experiment of ($R$-MBA)$_2$PbI$_4$ from linear-polarised (LP) pump and left-circular polarized probe (top row), linear-polarized probe (middle row) and right-circular-polarised probe (bottom row). Left column shows the Fourier-transformed residual TA signal and Right column shows the corresponding integrated signal along probe wavelength**.**

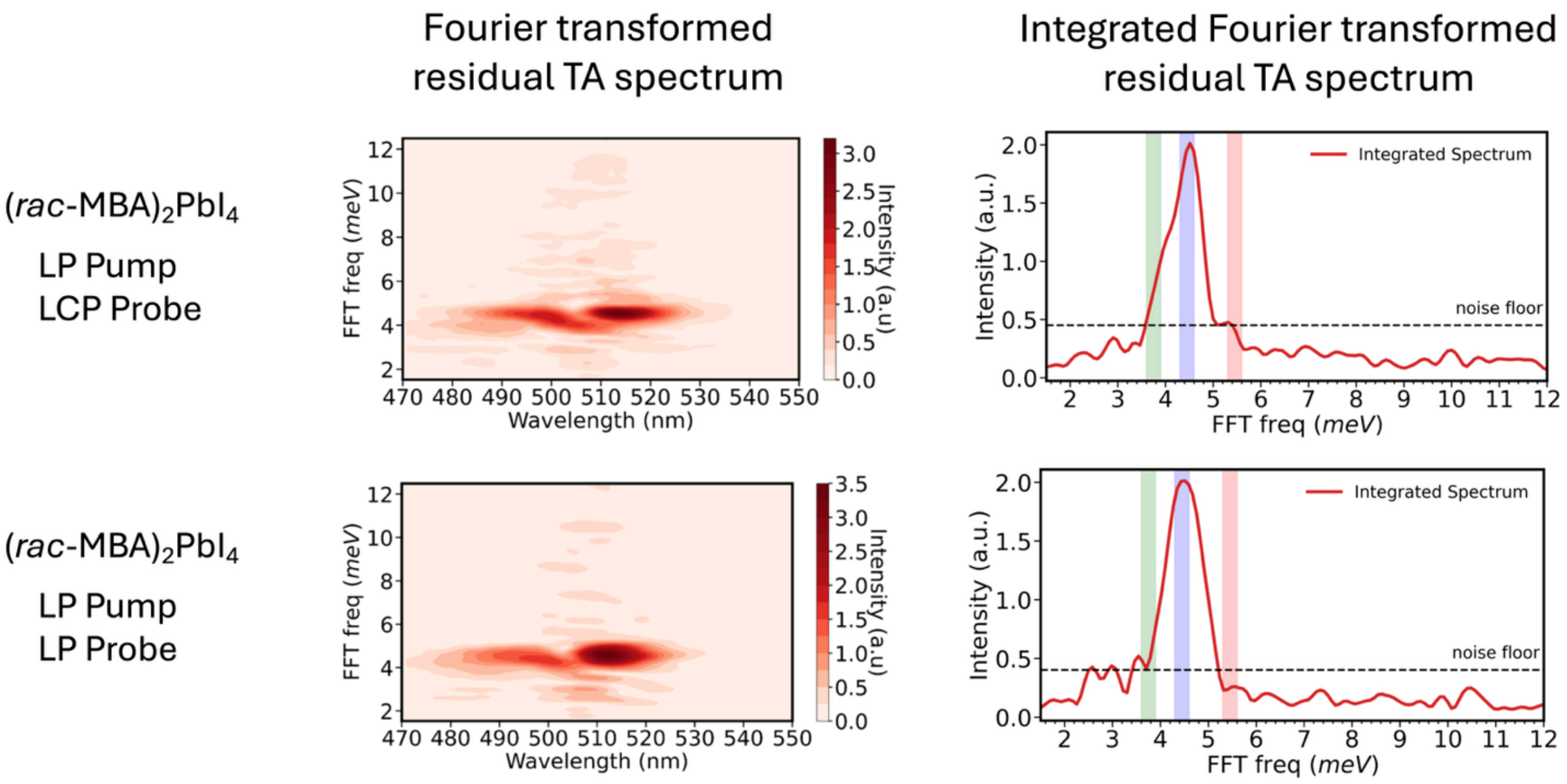

**Figure S7:** Residual oscillatory signal in the transient absorption experiment of (*rac*-MBA)$_2$PbI$_4$ from linear-polarised (LP) pump and left-circular polarized probe (top row) and linear-polarized probe (bottom row). Left column shows the Fourier-transformed residual TA signal and Right column shows the corresponding integrated signal along probe wavelength.

## 3. Transient Absorption Experiments at 77 K

Transient Absorption Experiment were conducted at 77 K with 400 nm, 8 μJ/cm$^2$, linear polarized pump pulses and with probe pulses in Left Circular (LCP) and Right Circular (RCP) polarized configurations to probe the transient circular dichroism signal from the sample. Below are the collected datasets, shown up to 12 ps in different selected probe configurations.

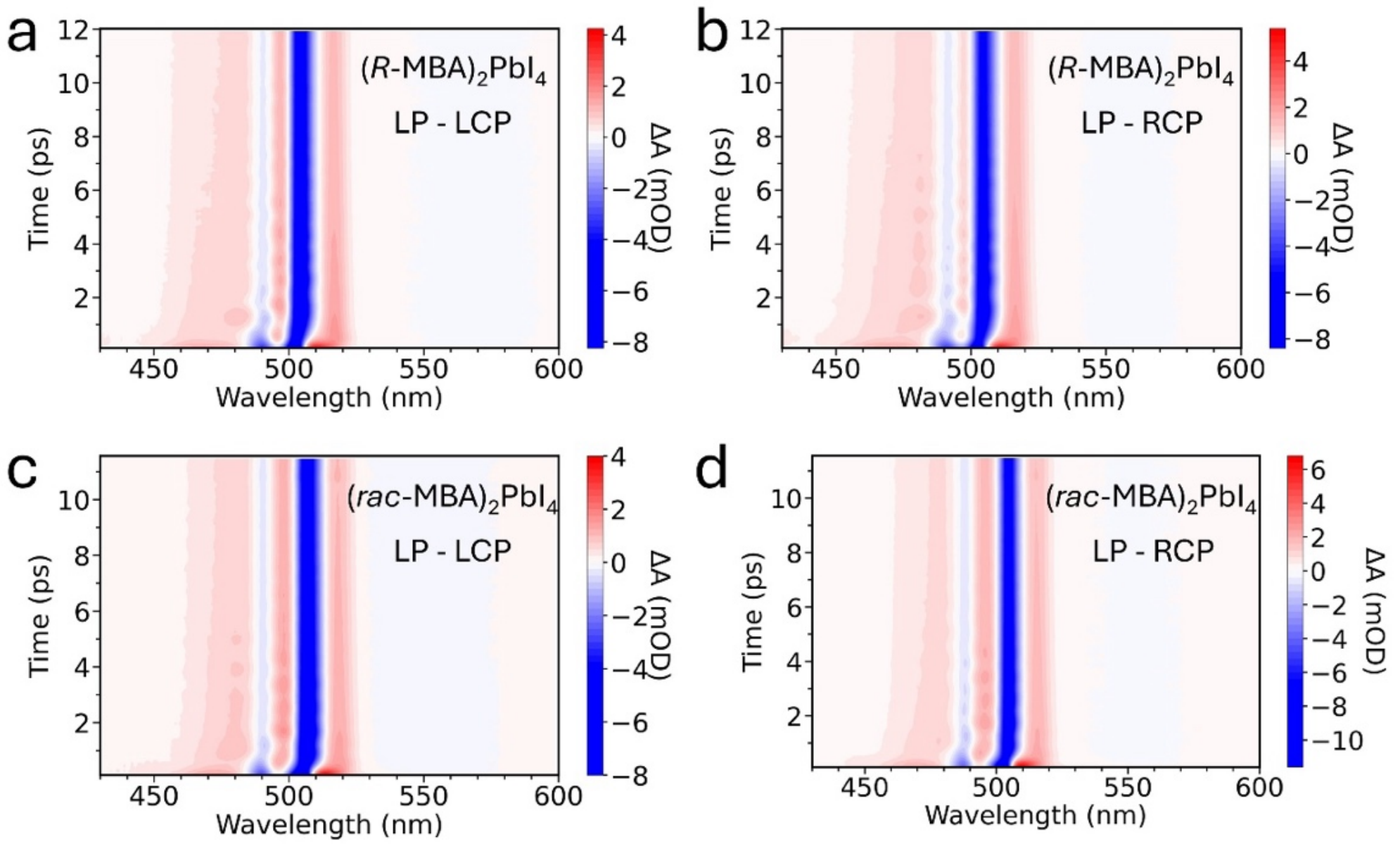

**Figure S8**: Pseudo-colour transient absorption colormap from 0 to 12 ps using linearly-polarized pump pulses, of (*R*-MBA)$_2$PbI$_4$ with left circularly-polarized probe pulses (a) and right circularly polarized pulses (b); of (*rac*-MBA)$_2$PbI$_4$ with left circularly-polarized probe pulses (c) and right circularly polarized pulses (d).

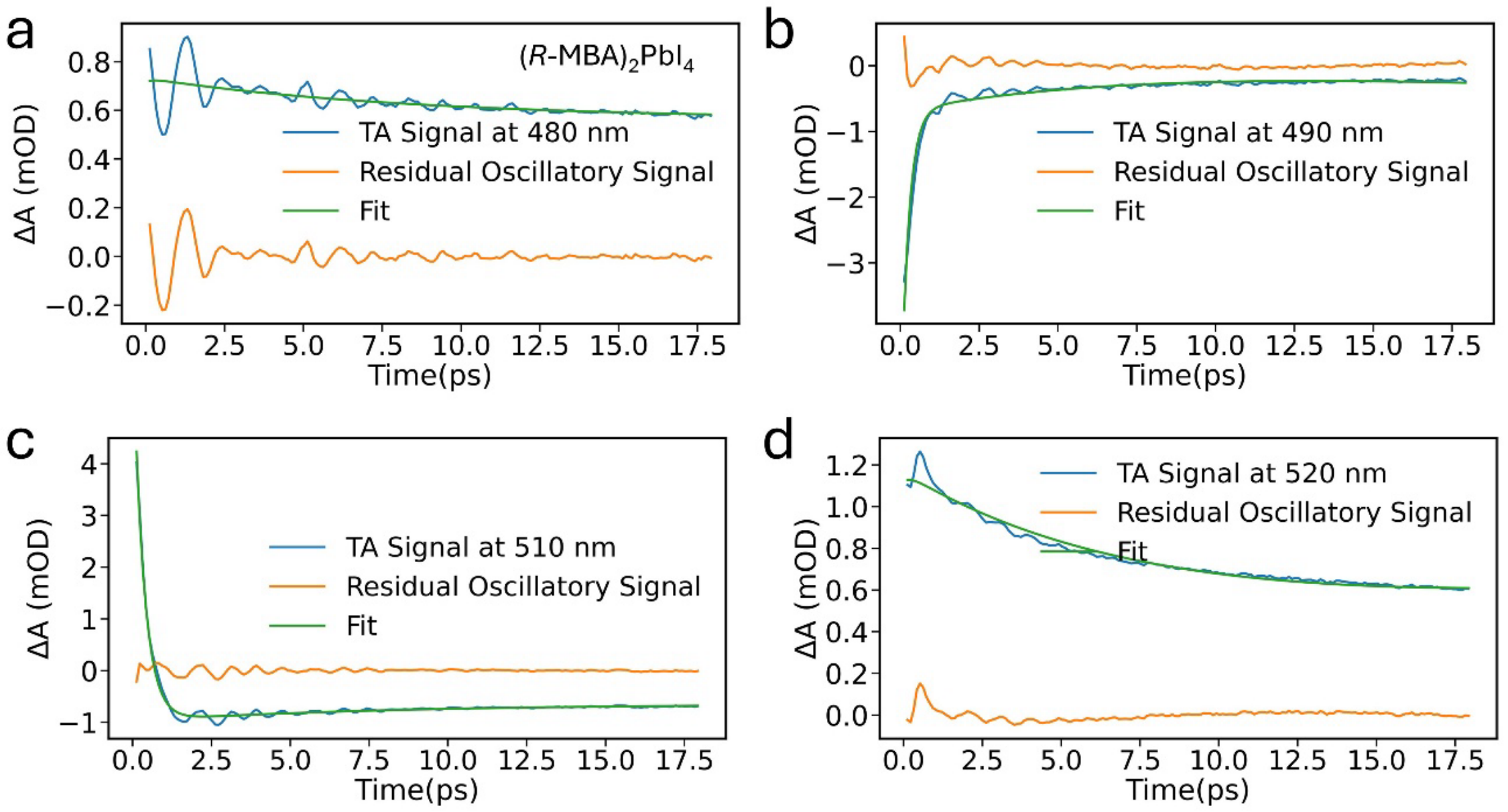

**Figure S9**: Calculation of Residual signal from transient absorption spectra at probe wavelengths for (R-MBA)$_2$PbI$_4$ , measured in the LP-LCP configuration at 298 K: (a) 480 nm (b) 490 nm (c) 510 nm and (d) 520 nm.

### 3.1 Effect of Changing Probe Polarization on Coherent Phonon Signals

In both ($R$-MBA)$_2$PbI$_4$ and ($rac$-MBA)$_2$PbI$_4$ thin films, the vibrational frequencies do not show considerable shift depending on probe pulse polarization. Frequency resolution of the Fourier transformed spectrum is ~0.3 meV.

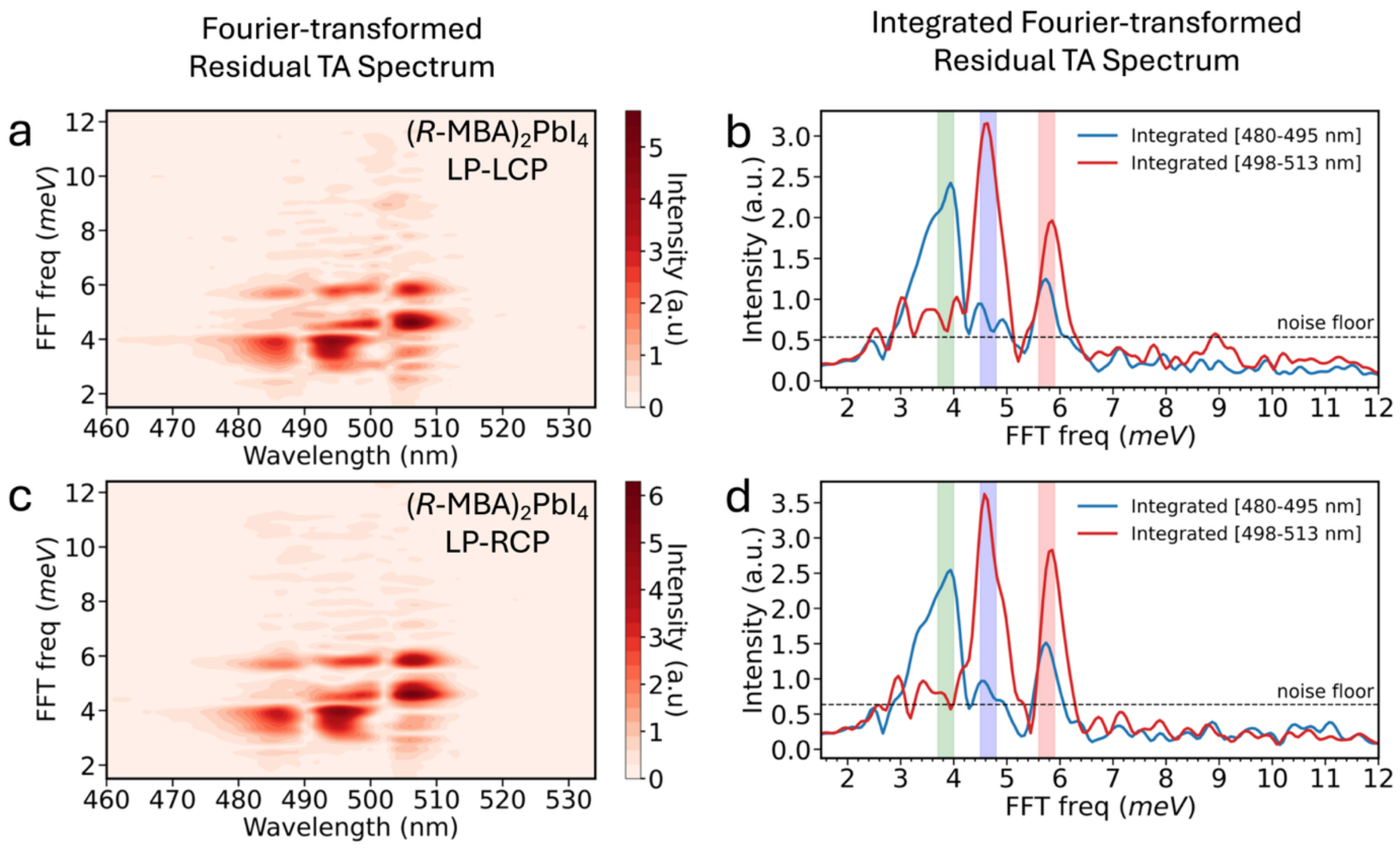

**Figure S10:** Residual oscillatory signal in the transient absorption experiment of ($R$-MBA)$_2$PbI$_4$ from linear-polarised (LP) pump and **(a)** left- and **(c)** right-circular-polarised (LCP/RCP) probe. **(b)** and **(d)** are the Fourier-transformed frequency map of oscillations of **(a)** and **(c)**, respectively.

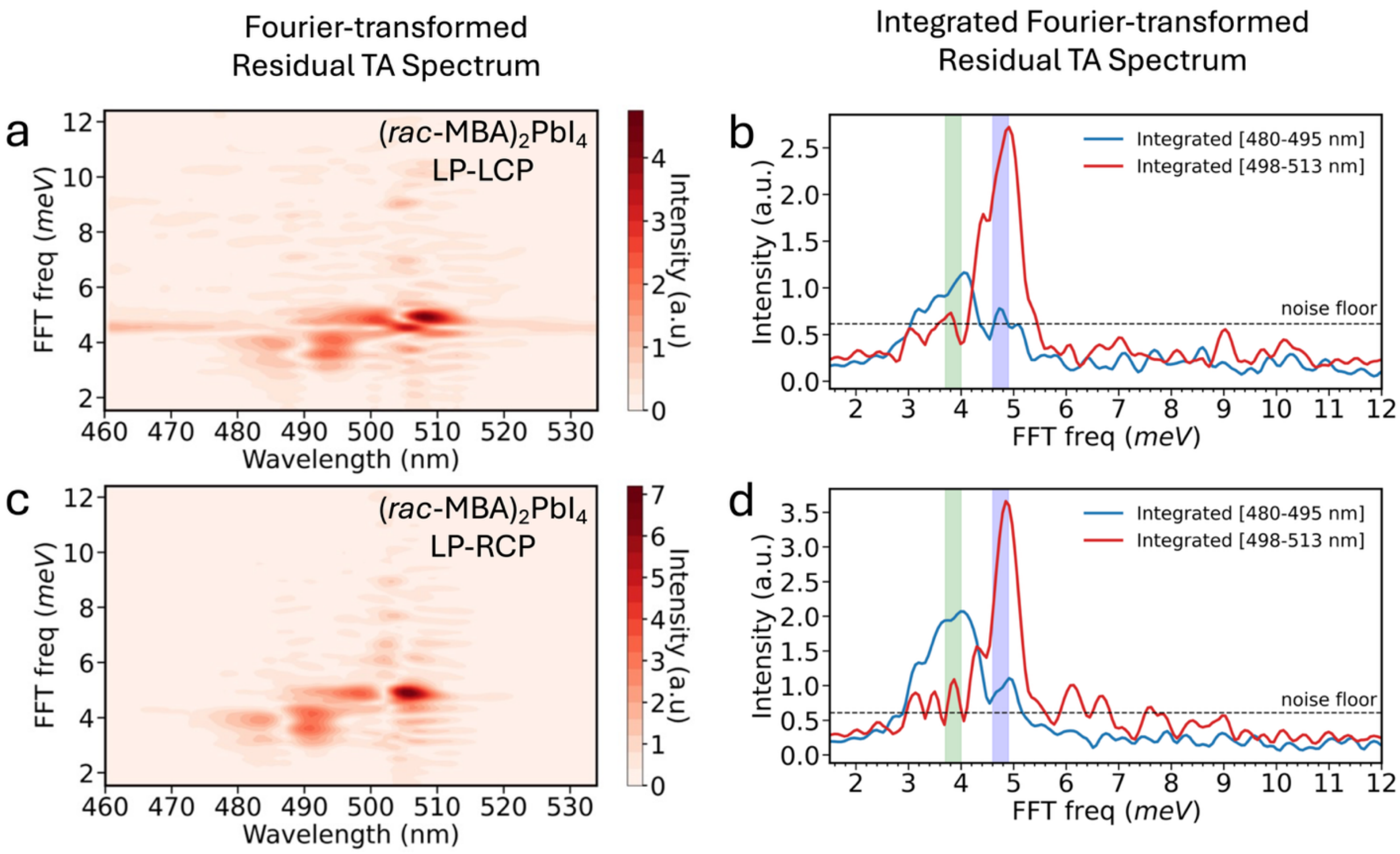

**Figure S11**: Residual oscillatory signal in the transient absorption experiment of (*rac*-MBA)$_2$PbI$_4$ from linear-polarised (LP) pump and **(a)** left- and **(c)** right-circular-polarised (LCP/RCP) probe. **(b)** and **(d)** are the Fourier-transformed frequency map of oscillations of **(a)** and **(c)**, respectively.

## 4. Detection Threshold in Integrated Raman Spectra

To enable statistically rigorous comparison of phonon modes between samples and across experimental conditions, we define a detection threshold for all integrated FFT spectra presented in this work. The threshold is defined as three times the root-mean-squared (RMS) noise of the integrated FFT spectrum evaluated in the 10–12 meV spectral window, where no Raman-active phonon modes are expected based on both experimental observation and DFT calculations. This window is sufficiently far from the modes of interest (2–8 meV) that it is unaffected by the tails of any phonon peaks, while remaining within the same frequency axis as the features under discussion and therefore subject to the same measurement conditions and noise characteristics.

Formally, for an integrated FFT spectrum $I(\nu)$, the noise floor $I_{th}$ is defined as:

$$I_{th} = 3\sqrt{\frac{1}{N}\sum_{i}(I(\nu_i) - \langle I \rangle)^2}, \quad \nu_i \in [10{,}12]\ meV \qquad (S1)$$

where the sum runs over all $N$ frequency points in the 10-12 meV window and $\langle I \rangle$ is their mean. A spectral feature is considered confidently detected only when its peak amplitude exceeds $I_{th}$.

This threshold is computed independently for each integrated FFT spectrum, since the noise level may vary between measurements due to differences in signal amplitude, sample quality, and integration time. The threshold is shown as a horizontal dashed line in all integrated FFT spectra in the main text and Supplementary Information.

Application of this threshold to the key comparisons in this work yields the following conclusions. At room temperature, the integrated FFT spectrum of ($R$-MBA)$_2$PbI$_4$ shows three peaks confidently above the threshold (M1, M2 and M3, refer to Table 1 of Main Text). For ($rac$-MBA)$_2$PbI$_4$ at room temperature, only the M2 peak exceeds the threshold, while the M3 feature is not confidently detected in the racemic sample. At 77K, the M3 peak in ($R$-MBA)$_2$PbI$_4$ exceeds the threshold in both exciton A and B integration windows, while no equivalent feature exceeds the threshold in the equivalent ($rac$-MBA)$_2$PbI$_4$ spectra. We therefore use the term "not observed" rather than "absent" when referring to experimental spectra, reserving "absent" for the conclusion supported jointly by the experimental threshold analysis and the DFT calculations showing no Raman-active mode with rotational displacement character in the racemic structure in this energy range.

## 5. FFT Spectrum Resolution and Stability of Peaks

The frequency resolution of the Fourier transform analysis is determined by the total time window $T_{max}$ of the oscillatory residuals used as input:

$$\delta E_{res} = \frac{h}{T_{max}} \qquad (S2)$$

In our experiments, the oscillatory residuals are analyzed over a time window of $T_{max} \approx 15\ \mathrm{ps}$, giving $\delta E_{res} \approx\ 0.3\ \mathrm{meV}$.

**Figure S6**-**Figure S7** (298 K) and **Figure S10**-**Figure S11** (77 K) show the integrated FFT spectra of (*R*-MBA)$_2$PbI$_4$ and (*rac*-MBA)$_2$PbI$_4$ obtained with LCP, LP, and RCP probe pulses. In all cases, no systematic shift of peak positions exceeding the FT grid resolution of 0.3 meV is observed between probe configurations. We interpret this as evidence of the stability of the phonon mode frequencies under different probe helicities, that is, the phonon frequencies are robust to the probe polarization state. We explicitly note that this observation should not be interpreted as a definitive proof of chiral specificity or lack thereof: the 0.3 meV FT resolution is insufficient to resolve subtle frequency shifts or mode splitting that may be present at smaller scales. Furthermore, any helicity-dependent frequency splitting arising from phonon angular momentum in our system is expected to be sub-0.3 meV based on typical phonon angular momenta in related systems,[3] which falls below our detection limit. The relation of M3 mode (Table 1, Main Text) to chiral optical response is therefore established through the enantiomeric sign reversal of the transient CD signal $\Delta\Delta A$ between (*R*-MBA)$_2$PbI$_4$ and (*S*-MBA)$_2$PbI$_4$, which is independent of frequency resolution (Figure 5, Main text and Supporting Information Section 10).

We also note a weak but consistent intensity asymmetry between LCP and RCP probe configurations for the M3 mode in (*R*-MBA)$_2$PbI$_4$, visible in Figure S9: the RCP probe yields a slightly larger integrated intensity for this mode. While subtle and below the level of the noise floor for definitive quantification, this asymmetry is in the direction expected for a mode with helicity-sensitive coupling and is consistent with the transient CD observation of Figure 5. No such asymmetry is observed for the common M2 mode, which serves as an internal control.

## 6. Temporal Dephasing of Coherent Phonons: 298 K and 77 K

The decay of the coherent vibrations was investigated for a comparison of the temporal aspects of vibronic coupling between enantiomer and racemic samples. The oscillatory dynamics at the probe wavelength ~512 nm, corresponding to exciton B, measured at 298 K were extracted from the data as shown in **Figure S4** and fitted with the following equation:

$$y(t) = \sum_{n=1}^{4} a_n * \exp\left(-\frac{t}{t_n}\right) * \cos(2\pi v_n t + \phi_n) \qquad (S3)$$

Where $a_n, t_n, v_n, \phi_n$ are the amplitude, dephasing time, frequency, and phase of the $n^{th}$ oscillating component in the signal. The temporal analysis further confirms the involved phonon modes and provides information about the decay time of the vibrations.[9] The results of the fits are given in **Table S3**. The coherent phonon signal and fits of (*R*-MBA)$_2$PbI$_4$ and (*rac*-MBA)$_2$PbI$_4$ are shown in **Figure S12a** and **Figure S12c**, respectively. In these plots, the fitted red curve can be decomposed into sum of the components shown in **Figure S12b** and **Figure S12d,** respectively.

Table S3: Results of the fitting of dephasing dynamics at GSB for enantiomer (top row) and racemic sample (bottom row) at 298 K.

| Mode | (*R*-MBA)$_2$PbI$_4$<br>(*rac*-MBA)$_2$PbI$_4$ | | | |
|---|---|---|---|---|
| | Energy (meV) | Amplitude (mOD) | Dephasing time (ps) | Phase (π rad) |
| 1 | 5.5±0.1<br>- | 0.04±0.01<br>- | 3.0±1.0<br>- | 1.7±0.1<br>- |
| 2 | 4.4±0.1<br>4.4±0.1 | -0.20±0.03<br>-0.65±0.16 | 2.2±0.3<br>1.1±0.1 | 0.9±0.1<br>1.0±0.1 |
| 3 | 3.1±0.1<br>3.7±0.2 | 0.37±0.10<br>0.90±0.15 | 0.5±0.1<br>0.4±0.1 | 1.2±0.1<br>1.2±0.1 |

The decay of (*R*-MBA)$_2$PbI$_4$ (**Figure S12a**) can be constructed by summing three contributing modes at 5.5 meV, 4.4 meV, and 3.1 meV. The three modes correspond closely to the ones identified in the frequency analysis of coherent phonon modes. The decay of (*rac*-MBA)$_2$PbI$_4$ (**Figure S12c**) is reproduced by just two modes at 4.4 meV and 3.7 meV, and is relatively faster than of (*R*-MBA)$_2$PbI$_4$. The mode at 5.5 meV (mode 1 in **Table S3**) is only observed in the chiral sample from both the frequency and the time decay analyses.

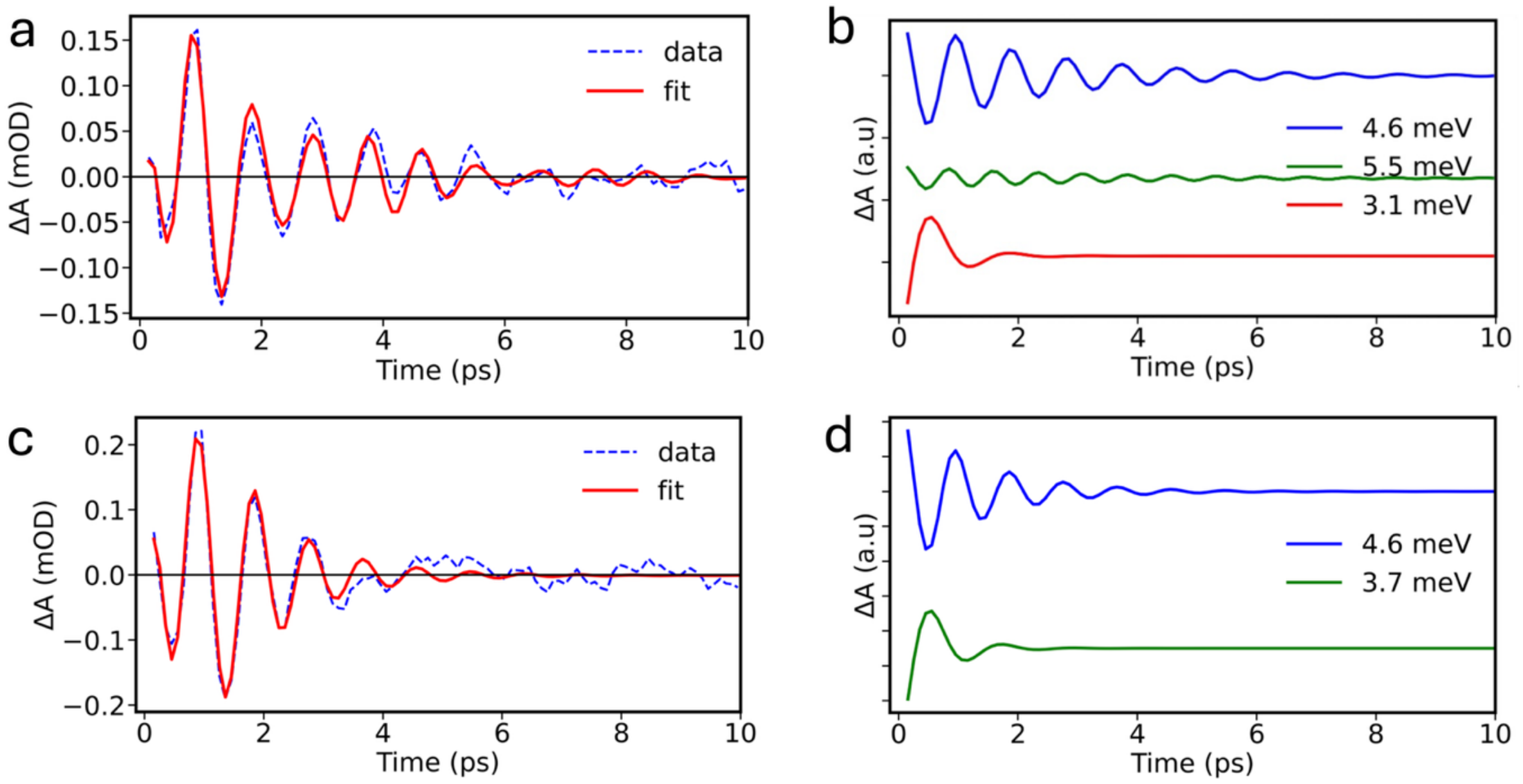

**Figure S12**: Comparison of vibrational dephasing in (*R*-MBA)$_2$PbI$_4$ (a-b) and (*rac*-MBA)$_2$PbI$_4$ (c-d) measured with linear-polarized (LP) pump and left circularly-polarized (LCP) probe at room temperature. (a) Residual oscillatory signal of (*R*-MBA)$_2$PbI$_4$ at 510 nm (blue dotted line) and Fit (red solid line). (b) Four components of the fit in (a). (c) Residual oscillatory signal of (*rac*-MBA)$_2$PbI$_4$ at 510 nm (blue dotted line) and Fit (red solid line). (d) Two components of the fit in (c). It can be concluded that a relatively long-lived (~ 3ps) M3 coherent phonon is absent in the racemic mixture.

The temporal analysis also indicates the mechanism of generation of coherent phonons, from the initial phases $\phi_n$ reported in **Table S3**. A cosine-like phase indicates the displacive excitation of coherent phonons (DECP). In DECP, photoexcitation shifts the potential energy surface minimum of the excited state along the phonon coordinate, and the system begins to oscillate from the ground-state geometry toward the new equilibrium. This produces a cosine-like oscillation whose amplitude is proportional to the displacement of the excited-state potential minimum relative to the ground state. The DECP mechanism is expected when the pump pulse creates carrier populations that couple to the phonon coordinate, consistent with our above-gap excitation at 400 nm creating excitations in the electronic continuum that relax to strongly bound excitonic states. This contrasts with impulsive stimulated Raman scattering (ISRS), which produces sine-like phases and operates through a virtual Raman process without creating real carrier populations. From **Table S4**, the 5.5 meV mode has a sine-like phase indicating origin from the ISRS process, while the rest of the modes have a cosine-like phase which implies that they are generated via the DECP process.[9,10]

The oscillatory dynamics at the probe wavelength ~506 nm, measured at 77 K is shown below in **Figure S13** and **Table S4** for understanding the effect of temperature. The mode frequencies, amplitude, dephasing times and initial phase are obtained from fitting the dynamics with equation $S3$. For both samples, the modes in the 3.5–5.5 meV range show a blueshift on cooling from 298 K to 77 K, on the order of 0.1–0.3 meV, consistent with the standard anharmonic contraction of the lattice and reduction of thermal expansion-driven softening expected for optical phonons in this energy range.

The dephasing times are longer at 77 K, consistent with the reduced availability of thermal modes to couple to the coherent vibrations. For (*R*-MBA)$_2$PbI$_4$ an additional mode at ~3.0 meV is found to contribute to the decay. This mode may have a strong temperature dependent dephasing, hence its decay may be too fast to be detected at 298 K. As in the room temperature scenario, the initial phases are mostly cosine-like (DECP), except for the 5.7±0.1 meV mode which has a sine-like phase consistent with the ISRS-driven excitation of coherent modes. The temporal analysis helps to further clarify the mode frequencies and dynamical aspects of the generation and decay.

Table S4: Results of the fitting of dephasing dynamics at GSB of exciton-B for enantiomer (top row) and racemic sample (bottom row) at 77 K.

| Mode | (*R*-MBA)$_2$PbI$_4$<br>(*rac*-MBA)$_2$PbI$_4$ | | | |
|---|---|---|---|---|
| | Energy (meV) | Amplitude (mOD) | Dephasing time (ps) | Phase (π rad) |
| 1 | 5.7±0.1<br>- | -0.21±0.04<br>- | 2.9±0.5<br>- | 1.4±0.1<br>- |
| 2 | 4.6±0.1<br>4.9±0.1 | -0.42±0.05<br>-0.49±0.06 | 3.1±0.4<br>2.3±0.2 | 1.2±0.1<br>1.0±0.1 |
| 3 | 3.5±0.1<br>3.6±0.2 | 1.60±0.30<br>0.31±0.06 | 0.6±0.1<br>1.1±0.2 | 1.2±0.1<br>1.2±0.1 |
| 4 | 3.0±0.1<br>- | -0.88±0.19<br>- | 1.3±0.2<br>- | 1.2±0.1<br>- |

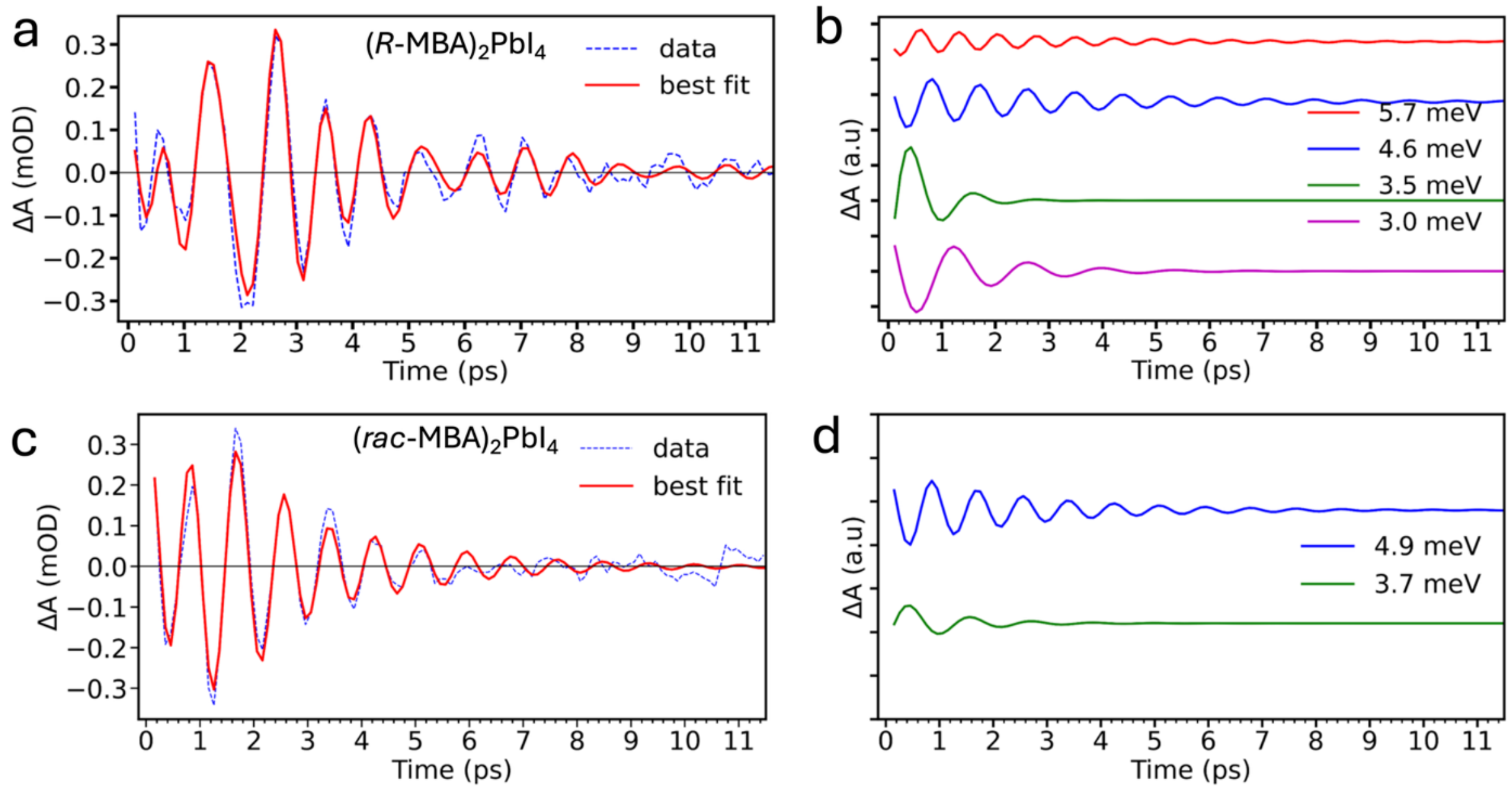

**Figure S13:** Comparison of vibrational dephasing in $(R\text{-MBA})_2PbI_4$ (a-b) and $(rac\text{-MBA})_2PbI_4$ (c-d) measured with linear-polarized (LP) pump and left circularly-polarized (LCP) probe at 77 K. (a) Residual oscillatory signal of $(R\text{-MBA})_2PbI_4$ at 506 nm (blue dotted line) and Fit (red solid line). (b) Four components of the fit in (a). (c) Residual oscillatory signal of $(rac\text{-MBA})_2PbI_4$ at 507 nm (blue dotted line) and Fit (red solid line). (d) Two components of the fit in (c). It can be concluded that a relatively long-lived (~ 3ps) M3 coherent phonon is absent in the racemic mixture.

## 7. Discussion on difference in vibrational coupling of exciton A and B between ($R$-MBA)$_2$PbI$_4$ and ($rac$-MBA)$_2$PbI$_4$

At 77 K, the transient absorption spectra of both ($R$-MBA)$_2$PbI$_4$ and ($rac$-MBA)$_2$PbI$_4$ show two distinct ground state bleach (GSB) features at ~490 nm and ~502 nm, which we label exciton A and exciton B in order of increasing energy. The energy separation of approximately 60 meV between these features rules out a phonon replica origin, since the relevant phonon modes in the inorganic Pb-I framework have energies of 2–8 meV, far below the observed splitting. The two states are instead attributed to distinct excitonic species with different polaronic character, consistent with previous reports of coexisting excitonic states in 2D lead halide perovskites.[4,5] The different dynamics of the two GSB features further confirm their distinct character.

In the chiral sample, integration of the FT residual maps over the exciton A (480-495 nm) and exciton B (498-513 nm) spectral windows reveals that the two excitons couple to the lattice with comparable overall strength but with mode-selective differences. Exciton A couples preferentially to modes at 3.7 and 5.7 meV, while exciton B couples preferentially to modes at 4.6 and 5.7 meV. The 5.7 meV mode therefore couples to both excitonic states, while the 3.7 and 4.6 meV modes show selectivity for exciton A and B respectively. This mode-selective coupling pattern is consistent with the polaronic picture: excitons A and B represent distinct bound states of the electron-hole pair with different lattice deformation patterns, and each couples most strongly to the phonon mode that best matches its own deformation geometry.

In the racemic sample, the coupling landscape is qualitatively different. Exciton A couples only weakly to lattice vibrations, with contributions from modes below 4 meV that are near or at the detection threshold. Exciton B couples predominantly to a single mode at 4.9 meV, with substantially stronger coupling than exciton B shows to any single mode in the chiral sample. The 5.7 meV mode is not observed from the racemic spectrum in both integration windows.

The stronger coupling of exciton B to the lattice in ($rac$-MBA)$_2$PbI$_4$, relative to the more comparable A/B coupling in ($R$-MBA)$_2$PbI$_4$, could reflect either (i) a difference in exciton-phonon coupling between the two space groups, or (ii) a difference in the orbital/inorganic character of excitons A and B between the two structures, without requiring the phonon landscape to differ.

The differential A/B coupling is consistent with the former scenario. This is supported by the From the phonon dispersion and Raman-activity calculations, the set of low-energy modes available to couple to either exciton is more for ($R$-MBA)$_2$PbI$_4$ compared to ($rac$-MBA)$_2$PbI$_4$, most notably the absence of a Raman-active mode near 5.7 meV with organic-inorganic coupled character in the racemic structure. This establishes that at least part of the observed difference in A/B coupling strength is attributable to a genuinely different phonon landscape, not solely to a reassignment of exciton character. We cannot, however, fully rule out a concurrent contribution from (ii). For example, the two structures may have different Pb–I–Pb bond angles and different local symmetry at the exciton site, which is known to modulate exciton orbital character and band-edge composition in 2D perovskites.[6] Rigorously separating the two contributions would require exciton-phonon coupling matrix elements computed explicitly for both structures. This would constitute a substantial many-body calculation that we defer to future work.

## 8. Vibrational Modes, Raman Activity: Theoretical Calculations

The chiral and racemic samples crystallize in different space groups ($P2_12_12_1$ and $P2_1/c$ respectively), so a vibrational mode could in principle exist in the racemic structure but be Raman-inactive due to symmetry selection rules rather than being physically absent. To address this and to clarify the nature of vibrational modes in both materials, we report full phonon dispersion calculations, Raman activity analysis and contribution of the organic and inorganic sublattices to each vibrational mode in the frequency range of interest.

The complete phonon dispersions of both structures are shown in

**Figure S14**. Both samples have a dense phonon spectrum in the 0-10 meV range, as typical of the low energy phonon spectrum of metal halide perovskites.[7] Pb and I atoms have a dominant contribution to the density of states, whereas the organic spacer has a minor contribution. The theoretical Raman spectra are shown in

**Figure S15.** The calculated modes are indicated by the blue stick spectra, where it can be observed that the chiral sample has more Raman-active modes within the 0-10 meV range. The spectra are obtained after incorporating broadening. The red sticks indicate the experimentally observed phonon modes from the TA experiments.

For $(R\text{-MBA})_2PbI_4$ four modes are identified, marked by the yellow stars on

**Figure S15**: 3.6 meV, 5.1 meV, 5.9 meV and 7.4 meV. Modes observed TA experiments are at 3.7 ± 0.2 meV, 4.4 ± 0.2 meV, and 5.4 ± 0.2 meV (M1, M2 and M3). By order of the energy, M1 is assigned to the 3.6 meV mode and M2 is assigned to the 5.1 meV mode This assignments are consistent by energy with an offset of a few meV. It is well established that the computational results systematically overestimate vibrational frequencies in hybrid perovskites due to underestimation of force constants;[8] the peak order and number are therefore a more reliable assignment criteria than absolute energies alone. Using the finite-difference Raman activities and incorporating a 7 $cm^{-1}$ broadening to simulate the realistic spectra, the 5.9 meV and 7.4 meV modes contribute additively to a single broadened spectral feature (

**Figure S15**). The experimental M3 peak is therefore assigned as the unresolved envelope of these closely spaced modes.

For $(rac\text{-MBA})_2PbI_4$, the calculation identifies a dominant Raman-active mode at ~5.2 meV, corresponding to the experimental M2 peak (4.4 ± 0.2 meV). Although a minor peak appears at ~7.9 meV, this peak was not conclusively detected in our experiments.

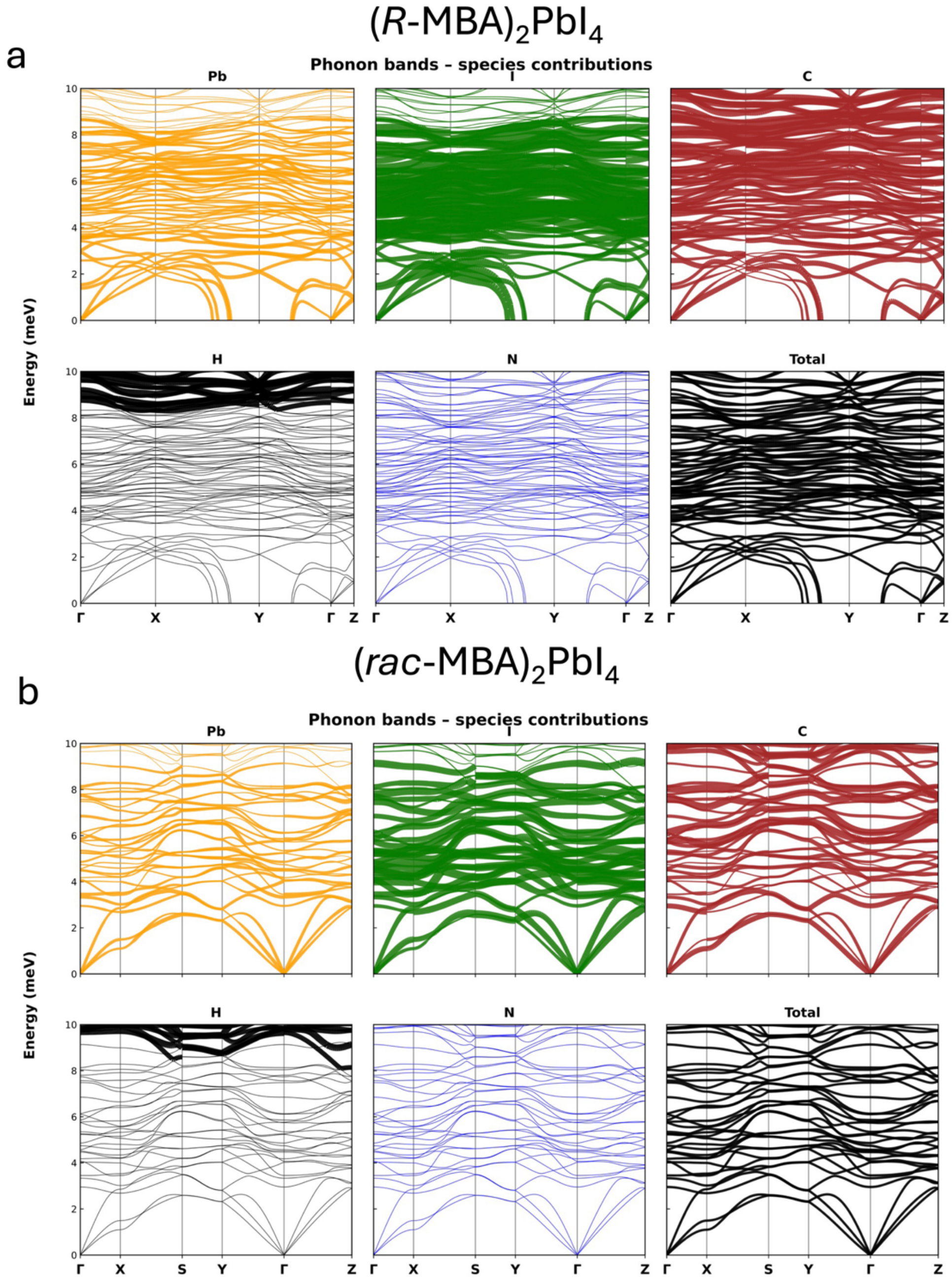

**Figure S14**: Phonon Dispersion in ($R$-MBA)$_2$PbI$_4$ (top) and ($rac$-MBA)$_2$PbI$_4$ (bottom) along with contribution from individual species represented by the different colours.

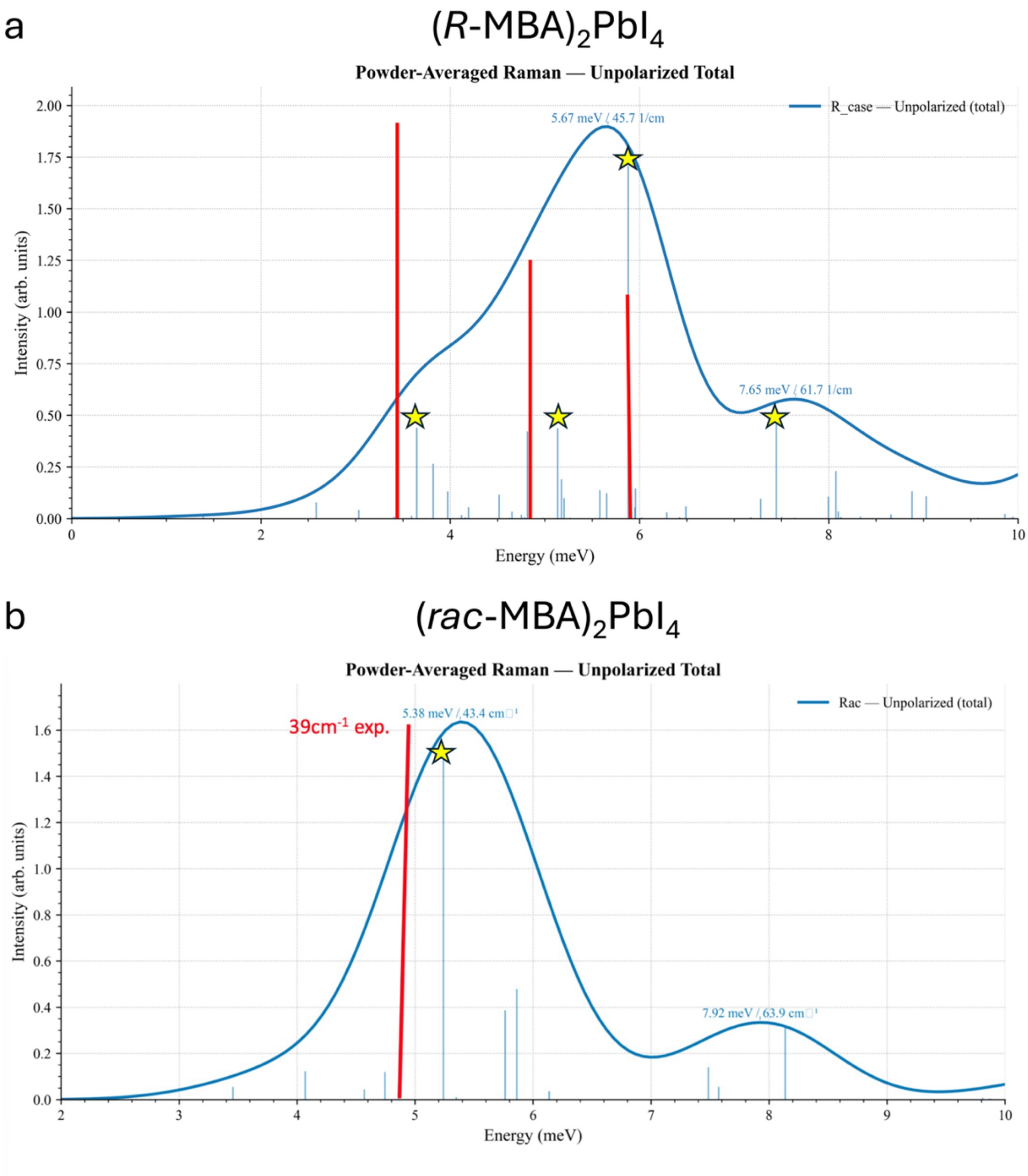

**Figure S15:** Computed Raman Spectra with Boltzmann correction applied and peaks broadened for ($R$-MBA)$_2$PbI$_4$ (top) and (*rac*-MBA)$_2$PbI$_4$ (bottom). Blue sticks indicate discrete modes. Red sticks indicate experimentally observed modes. The modes labelled with star analyzed with their vibrational displacement patterns.

As we can assert the atomic displacements with the phonon mode assignment in theory, we analysed all vibrational modes, in comparison to the modes just in the Pb-I sub-lattice (

**Figure S16**), enabling isolation of modes unique to the inorganic part of the lattice. For both samples, while modes in 3.5 meV to 5.7 meV range show predominantly vibrations in the Pb-I lattice, modes at energy > 5.7 meV show vibrations spanning the inorganic and organic octahedra, denoted by the larger contribution from carbon, ‘C’.

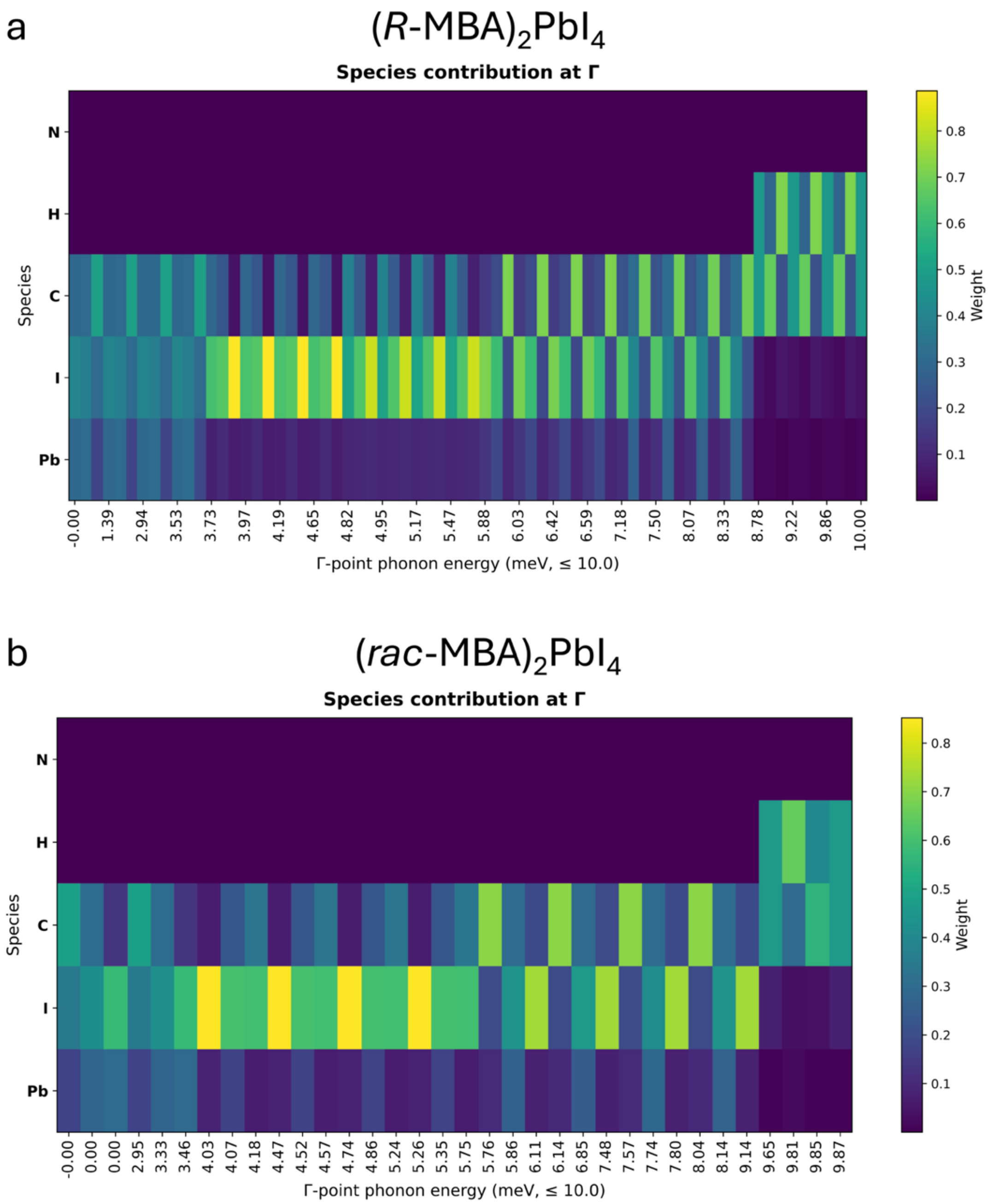

**Figure S16**: Correlation between phonon mode frequency at $\Gamma$-point and their species-wise decomposition: ($R$-MBA)$_2$PbI$_4$ (top) and ($rac$-MBA)$_2$PbI$_4$ (bottom).

Further calculation showed that modes in the energy range show the rotational displacement patterns in ($R$-MBA)$_2$PbI$_4$ only (Figure 4b, Main text). These modes are visualized with larger unit cell representations and rescaled displacement arrows to illustrate the rotational-vibrational coupling character of the mode (**Figure S17**-**Figure S20**). Focusing on the theoretically calculated modes marked by stars in amine groups of the organic spacer (**Figure S17**). This mode lacks significant contribution from the organic layer.

**Figure S15**, in ($R$-MBA)$_2$PbI$_4$ the dominant low-energy mode at 3.6 meV primarily involves linear displacements concentrated on the inorganic octahedra and the nearby

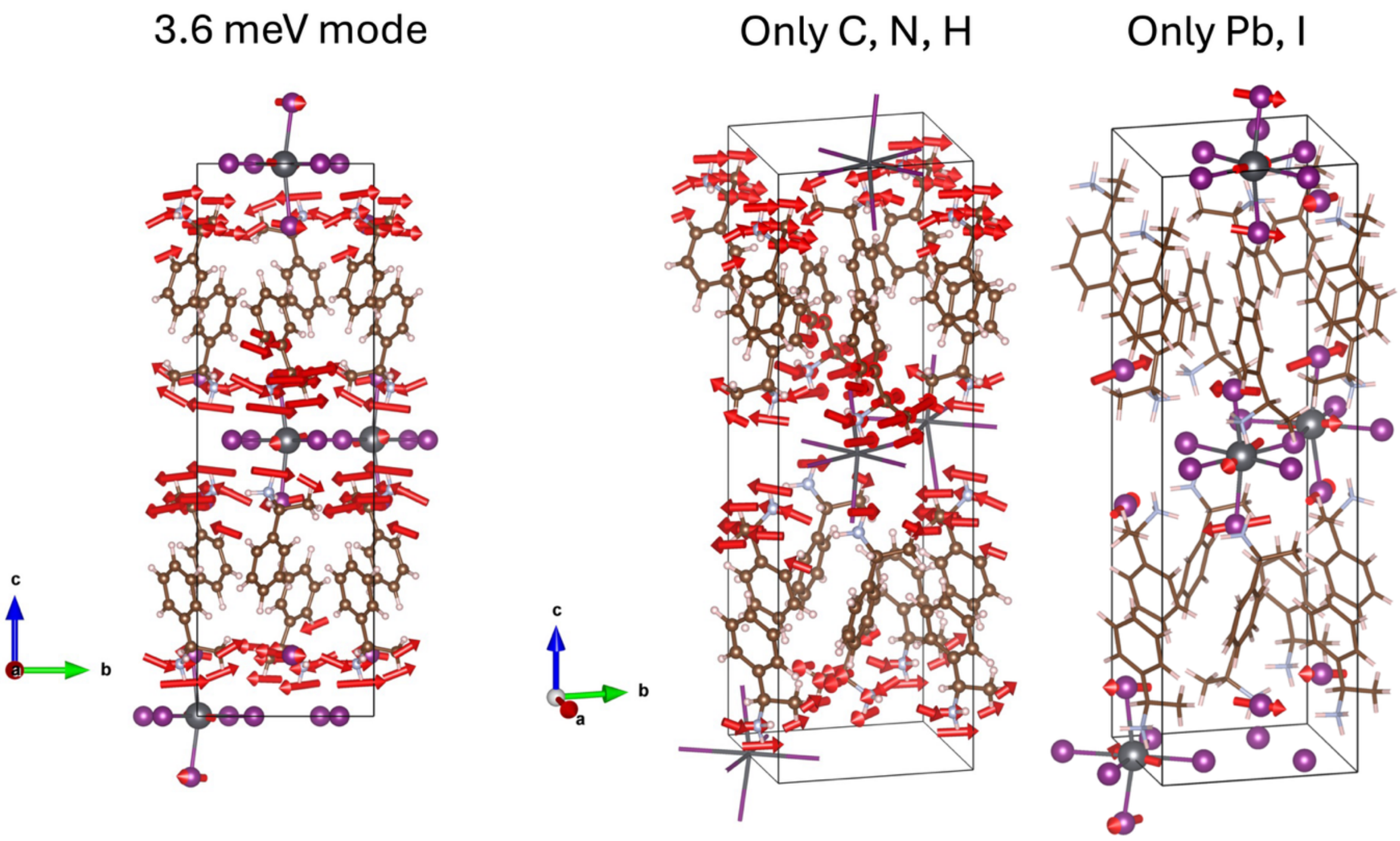

**Figure S17:** Representation of Raman-active vibrational mode at ~3.6 meV of (*R*-MBA)$_2$PbI$_4$ (left). Vibrational pattern in the organic and inorganic parts are shown separately in the middle and right.

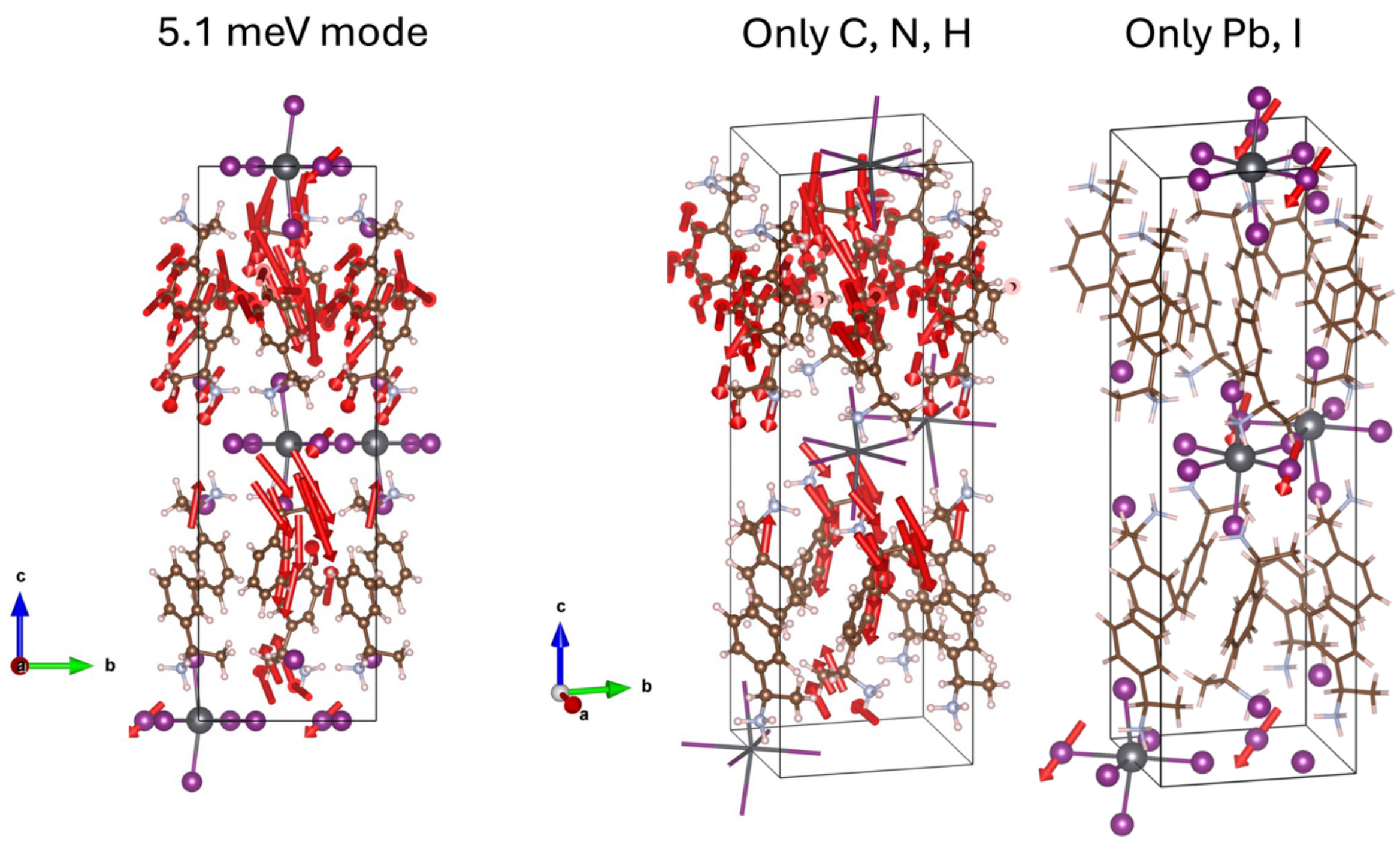

**Figure S18:** Representation of Raman-active vibrational mode at ~5.1 meV of (*R*-MBA)$_2$PbI$_4$ (left). Vibrational pattern in the organic and inorganic parts are shown separately in the middle and right.

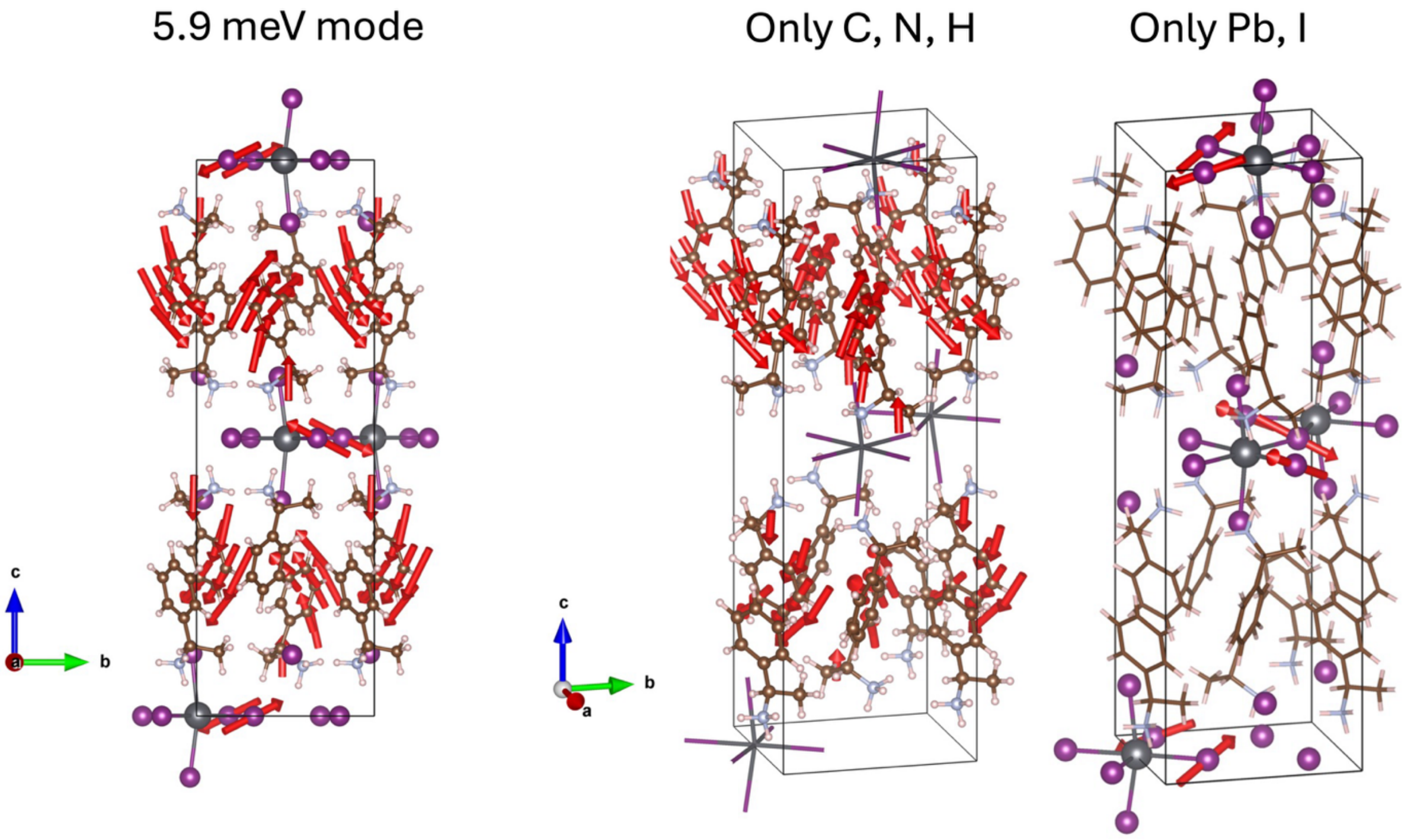

**Figure S19:** Representation of Raman-active vibrational mode at ~5.9 meV of (*R*-MBA)$_2$PbI$_4$ (left). Vibrational pattern in the organic and inorganic parts are shown separately in the middle and right.

The 5.1 (**Figure S18**), 5.9 (**Figure S19**) and 7.4 meV (**Figure S20**) display a substantially larger contribution from the organic framework with relatively lower contribution from the inorganic layer. Among these, the 5.9 meV mode visually represents a rotational displacement pattern within its organic layer. This mode which is unique to the chiral sample also matches energetically with the M3 mode that we spectroscopically observe in ($R$-MBA)$_2$PbI$_4$ and not in ($rac$-MBA)$_2$PbI$_4$.

Even though the mode shows the rotational displacement patterns in ($R$-MBA)$_2$PbI$_4$, They do not constitute a proof of phonon chirality itself. The experimental feature of our interest at ~5.7 meV in ($R$-MBA)$_2$PbI$_4$ likely reflects a coherent superposition of the 5.1, 5.9, and 7.4 meV modes rather than any one of them individually, enabled by their proximity and by mixing in the low-symmetry chiral structure.

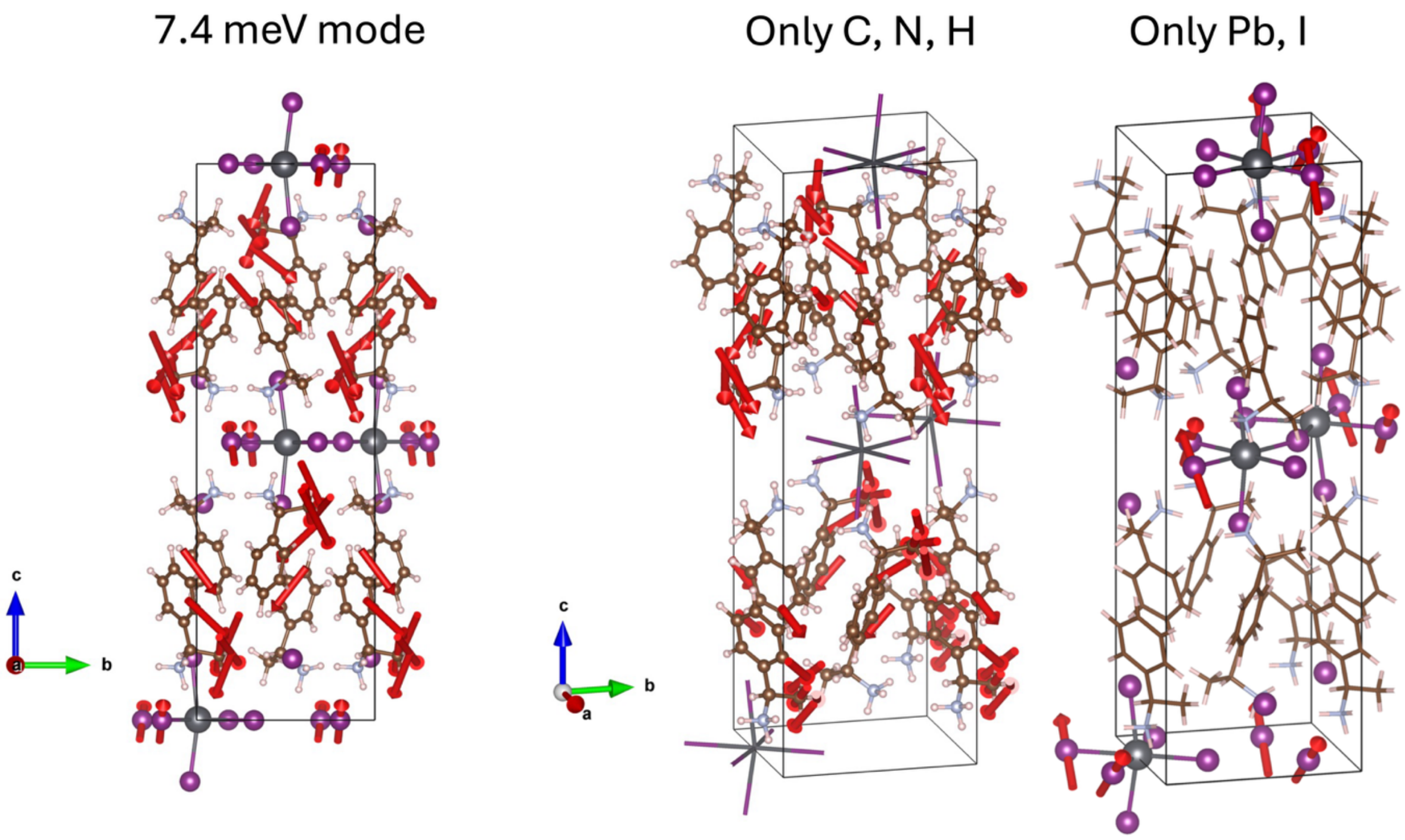

**Figure S20:** Representation of Raman-active vibrational mode at ~7.4 meV (left). Vibrational pattern in the organic and inorganic parts are shown separately in the middle and right.

## 9. Optical, Structural Properties and Ultrafast Dynamics in (*S*-MBA)$_2$PbI$_4$

We prepared the *S*-enantiomer (*S*-MBA)$_2$PbI$_4$ under identical conditions and performed systematic investigations to compare it to (*R*-MBA)$_2$PbI$_4$. The results and analysis are given below:

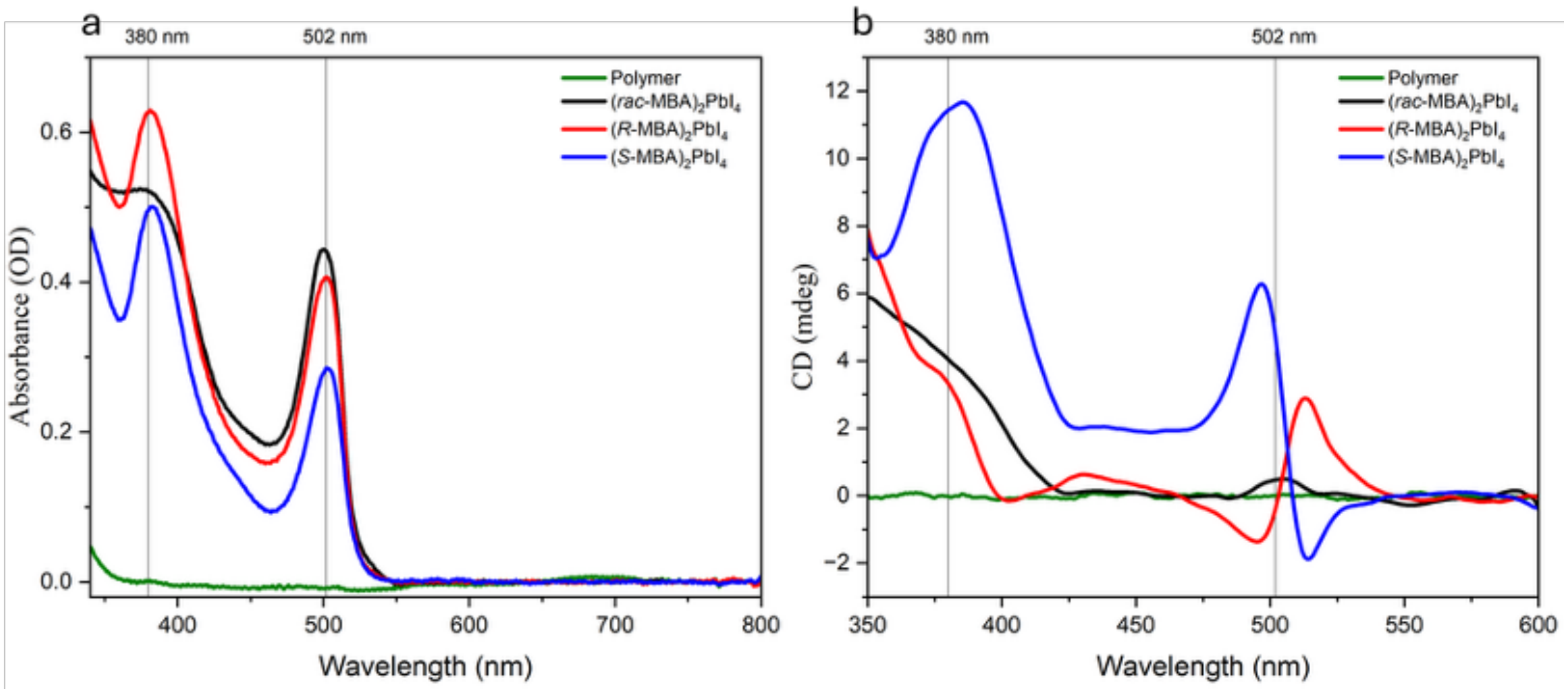

**Figure S21:** (a) UV–Vis absorption spectra showed an excitonic peak at ~502 nm in the three samples. Green spectra indicates absorption of the polymer used for encapsulating the perovskite film (b) CD spectra revealed a derivative-shaped peak for the chiral (R/S-MBA)2PbI4 film with opposite signs. No CD response was observed for the racemic sample and for the polymer at the band-edge.

**Figure S21** shows the room temperature UV-vis absorption and Circular dichroism (CD) spectra of the samples, showing that the films show similar absorption profiles while the CD of the enantiomers have their sign reversed, as expected. X-ray Diffraction (XRD) patterns of samples with each chiral enantiomer and with the racemic sample are shown in **Figure S22**.

Peaks at equal intervals in all samples originate from the (00*n*) planes for the 2D structure. The impurity peak at 8 degrees is absent for (*rac*-MBA)$_2$PbI$_4$, implying that there is very little of the 1D phase in it. Whereas for both the chiral samples, there appears a peak at 8 degrees originating from the 1D phase. In particular, the newly prepared (*S*-MBA)$_2$PbI$_4$ has a more prominent impurity peak in the XRD in comparison to the ($R$-MBA)$_2$PbI$_4$ sample. A recent study indicates that the ambient conditions including moisture during synthesis affect the phase composition.[1] Based on these reports, we report that we were unable to control the amount of impurity phase due to limitations in our synthesis procedure. Nevertheless, the spectroscopic measurement on ($S$-MBA)$_2$PbI$_4$ provided us important results regarding the role of the impurity 1D phase the chiral 2D lattice.

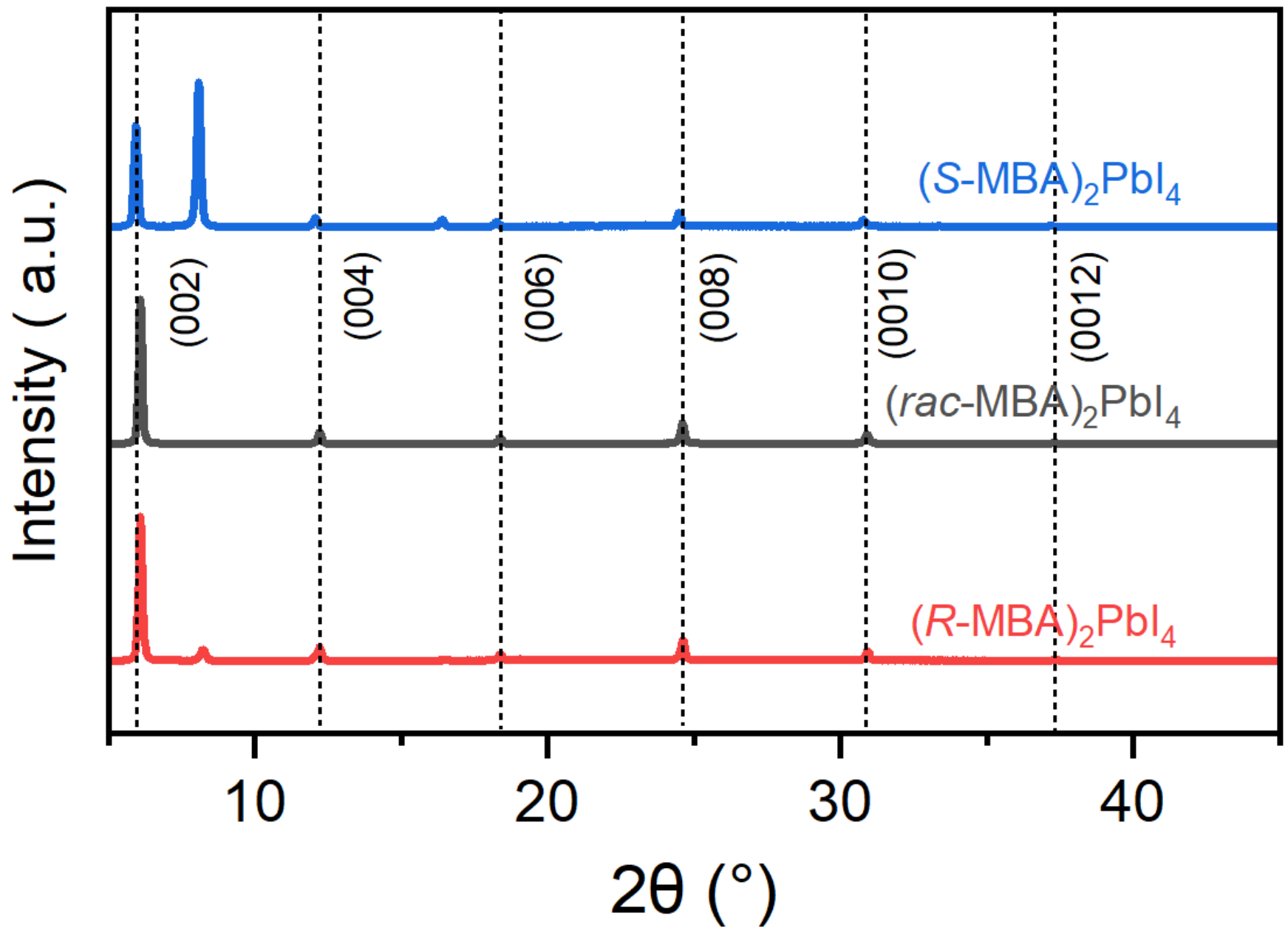

**Figure S22:** X-ray Diffraction of the thin film samples. Periodic peaks are observed, indicating their layered structure. The ($R$-MBA)$_2$PbI$_4$ shows an additional peak at 8 degrees that is reported to originate from a minority 1D phase.[1]

We proceeded to measure circular-polarized transient absorption spectra of the ($S$-MBA)$_2$PbI$_4$ film. Since XRD shows that ($S$-MBA)$_2$PbI$_4$ has a significantly larger concentration of the impurity phase, we hypothesized that if the ~5.7 meV is associated with the 1D phase, we would observe a higher intensity of that peak for ($S$-MBA)$_2$PbI$_4$ with respect to the other peaks. **Figure S23** shows the application of the residual TA signal analysis workflow applied to the ($S$-MBA)$_2$PbI$_4$ sample. The pump polarization was LP and the results are shown row-wise for each probe polarization (LCP, LP, RCP). It is clear that there is no enhancement of relative peak intensity of the 5.7 meV peak for the ($S$-MBA)$_2$PbI$_4$ sample. As discussed earlier, a higher proportion of impurity phase would have given rise to a stronger 5.7 meV peak relative to the other peaks. The observed similarity of the FFT spectra of both enantiomers indicates that the 5.7 meV has a common origin from the dominant 2D phase of the chiral samples.

We wish to also add that a recent study[3] reported the Raman spectrum of purely the 1D phase. By comparing their results to the present study, it becomes clear that several Raman peaks that are characteristic of the 1D phase are absent in our results. Based on this observation, we further confirm that the 5.7 meV mode in the present study originates from the dominant 2D phase with formula (*R/S*-MBA)$_2$PbI$_4$.

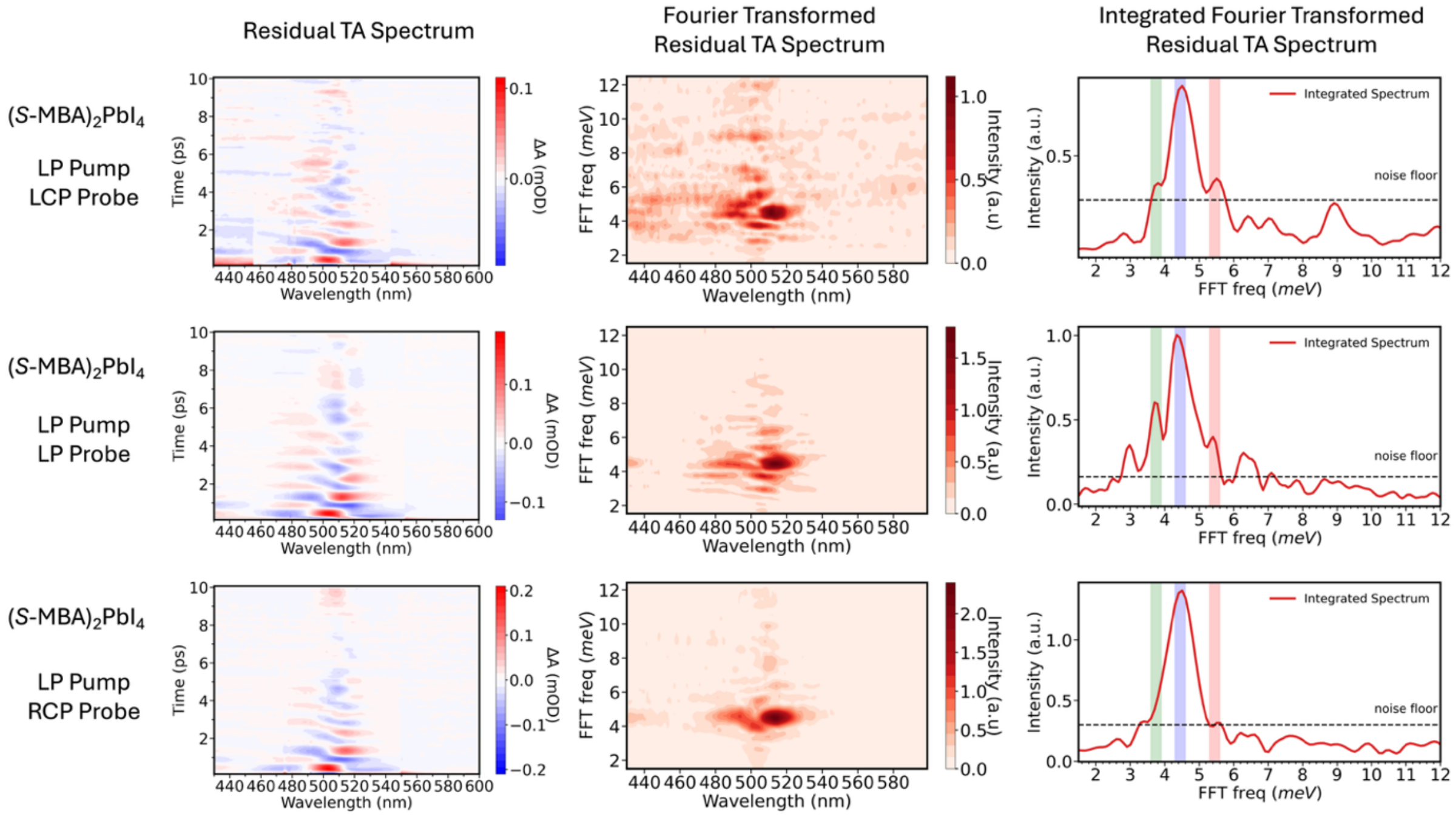

**Figure S23:** Residual TA Signal analysis of (*S*-MBA)$_2$PbI$_4$: Residual oscillatory signal in the transient absorption spectrum (left column), Fourier-transformed time-domain data (middle column) and Integrated signal intensity from probe wavelength 480 nm to 520 nm (right column), linear-polarized (LP) pump and left circularly-polarized (LCP) probe (top row), LP pump and LP probe (Middle row) and LP pump and RCP probe (bottom row).

## 10. Transient Circular Dichroism and Modulation of Rotational Strength

### A. Modulation of excitonic transition energy by coherent phonons

A coherent phonon is a lattice vibration in which the relevant unit cells oscillate in phase, so the vibrational coordinate follows a well-defined classical trajectory. Through deformation-potential coupling,[9] this vibration periodically shifts the exciton transition energy, which in turn makes the transient absorption (TA) signal oscillate at the phonon frequency. The normal-mode coordinate $Q_n$ corresponding to the mode $n$ has a well-defined trajectory:

$$Q_n(t) = Q_0 * \exp\left(-\frac{t}{t_n}\right) * \cos(2\pi\nu_n t + \phi_n) \qquad (S4)$$

Where $Q_0, t_n, \nu_n, \phi_n$ is the initial amplitude of the coherent displacement, coherence decay time, frequency of the mode and initial phase of oscillation. The exciton transition energy $E_{exc}$ depends on the lattice geometry. Expanding to first order in the displacement:

$$E_{exc}(Q_n) = E_0 + k_n Q_n \qquad (S5)$$

Where $E_0$ is the transition energy at equilibrium and $k_n$ is the deformation potential constant for the mode $n$, defined as

$$k_n = \frac{\partial E_{exc}}{\partial Q_n} \qquad (S6)$$

So the periodic energy shift due to the coherent lattice vibration is given by

$$\delta E(t) = E_{exc}(Q_n) - E_0 = k_n Q_0 \exp\left(-\frac{t}{t_n}\right) * \cos(2\pi\nu_n t + \phi_n) \qquad (S7)$$

Modelling the TA feature near the exciton transition as a fixed spectral shape whose centre tracks $E_{\text{exc}}$. When the centre moves by $\delta E(t)$, the whole shape slides by that amount without deforming:

$$\Delta A\big(\hbar\omega_{pr}, t\big) = \Delta A_0(\hbar\omega_{pr} - \delta E(t)) \qquad (S8)$$

Where $\Delta A$ is the instantaneous TA spectra, $\Delta A_0$ is the TA spectrum at equilibrium and $\hbar\omega_{pr}$ is the probe photon energy. Here we have assumed that the key effect on the spectra is modulation of the exciton transition energy and there is no modulation of amplitude of the transition. Because $\delta E$ is small compared with the peak width, a small horizontal slide produces a small vertical change set by the curve's slope. First-order Taylor expansion about $x = \hbar\omega_{pr}$:

$$\Delta A(\hbar\omega_{pr}, t) = \Delta A_0(\hbar\omega_{pr}) - \left[\frac{\partial \Delta A_0}{\partial(\hbar\omega_{pr})}\right] \delta E(t) \qquad (S9)$$

The first term on the right is the static transient absorption spectra, which is subtracted when the non-oscillatory part of the signal is removed via fitting. The second term gives the residual TA signal,

$$\Delta A^{res}(\hbar\omega_{pr}, t) = -\left[\frac{\partial \Delta A_0}{\partial(\hbar\omega_{pr})}\right] k_n * Q_0 \exp\left(-\frac{t}{t_n}\right) * \cos(2\pi\nu_n t + \phi_n) \qquad (S10)$$

The first term on the right $\left[\frac{\partial \Delta A_0}{\partial(\hbar\omega_{pr})}\right]$ is the derivative of the equilibrium TA spectrum, giving the dispersive lineshape about the ground state bleach peak position.

## B. Modulation of rotational strength of excitonic transition

Consider that the transition $0 \rightarrow a$ has the extinction coefficient $\varepsilon_{\pm}$ where $\pm$ denotes LCP/RCP:

$$\varepsilon_{\pm} \propto f(v). [|\mu_{0a}|^2 \pm 2\mathcal{R}_{0a}] \qquad (S11)$$

Where $f(v)$ is the normalized lineshape and $\mathcal{R}$ is the rotational strength $\mathcal{R}$ of a transition,[11,12]

$$\mathcal{R}_{0a} = Im[\mu_{0a}. m_{a0}] \qquad (S12)$$

$\mu, m$ are the electric and magnetic transition dipole moments. The absorption of linear polarized light $\propto (\varepsilon_+ + \varepsilon_-)$ and the circular dichroism $CD = A_{LCP} - A_{RCP} \propto (\varepsilon_+ - \varepsilon_-)$

$$(\varepsilon_+ + \varepsilon_-) \propto 2f(v). |\mu_{0a}|^2 \qquad (S13)$$

$$(\varepsilon_+ - \varepsilon_-) \propto 4f(v).\mathcal{R}_{0a} \tag{S14}$$

Suppose that due to exciton-phonon coupling, displacement along normal mode $Q_n$ modulates $\mathcal{R}$ (assuming the CD lineshape $\mathcal{L}(\omega)$ remains unchanged):

$$\mathcal{R}(t) = \mathcal{R}_0 + \frac{\partial \mathcal{R}}{\partial Q_n} Q_n(t) \tag{S15}$$

$$\mathcal{R}(t) = \mathcal{R}_0 + \frac{\partial \mathcal{R}}{\partial Q_n} Q_0 * \exp\left(-\frac{t}{t_n}\right) * \cos(2\pi v_n t + \phi_n) \tag{S16}$$

Where $\mathcal{R}_0 = \mathcal{R}_{0a}$, the steady-state rotational strength (Equation S12). The transient CD is defined as:

$$CD(t) - CD(0) = (A_{LCP} - A_{RCP})_{Pump\ ON} - (A_{LCP} - A_{RCP})_{Pump\ OFF} \tag{S17}$$

The CD signal at time $t$ is proportional to $\mathcal{R}(t)$ and is given by:

$$CD(t) = A_{LCP}(t) - A_{RCP}(t) \propto \mathcal{L}(\omega)\left[\mathcal{R}_0 + \frac{\partial \mathcal{R}}{\partial Q_n} Q_n(t)\right] \tag{S18}$$

the static CD is written as:

$$CD(0) = \mathcal{R}_0.\mathcal{L}(\omega) \tag{S19}$$

$\Delta\Delta A(t)$ is obtained in our study by subtracting the residual TA spectra: $\Delta A_{LCP}(t) - \Delta A_{RCP}(t)$.

$$\begin{aligned} \Delta A_{LCP}(t) - \Delta A_{RCP}(t) \\ = (A_{LCP}^{pump\ ON} - A_{LCP}^{pump\ OFF}) - (A_{RCP}^{pump\ ON} - A_{RCP}^{pump\ OFF}) \\ = (A_{LCP}^{pump\ ON} - A_{RCP}^{pump\ ON}) - (A_{LCP}^{pump\ OFF} - A_{RCP}^{pump\ OFF}) \\ = CD(t) - CD(0) \end{aligned} \tag{S20}$$

After the non-oscillatory features are subtracted from each differential absorption dataset. Therefore, the static CD term is removed, giving the final signal

$$\Delta\Delta A(t) = \frac{\partial \mathcal{R}}{\partial Q_n} Q_n(t).\mathcal{L}(\omega) \tag{S21}$$

$$\Delta\Delta A(t) = \frac{\partial \mathcal{R}}{\partial Q_n} . \mathcal{L}(\omega) . Q_0 \exp\left(-\frac{t}{t_n}\right) * \cos(2\pi\nu_n t + \phi_n) \qquad (S22)$$

Note that in the above steps $\Delta A$ refers to the residual TA spectrum, where non-oscillatory decay components are removed. Therefore, the obtained signal is the coherent phonon-induced transient CD. From Equation $S22$, It follows the original lineshape of the steady-state CD spectrum ($\mathcal{L}(\omega)$), while it is modulated at the frequency of the phonon coupled to the exciton ($\nu_n$). Importantly, $\mathcal{R}$ is a pseudo-scalar quantity that inverts sign under an improper rotation, therefore also the derivative term $\frac{\partial \mathcal{R}}{\partial Q_n}$ is pseudo-scalar. This implies that the sign of $\Delta\Delta A(t)$ will be inverted in (*S*-MBA)$_2$PbI$_4$ when compared with (*R*-MBA)$_2$PbI$_4$.

Since $\mathcal{R}$ is a pseudoscalar that changes sign under spatial inversion, the signal must therefore undergo an exact sign reversal between (*R*-MBA)$_2$PbI$_4$ and (*S*-MBA)$_2$PbI$_4$. $\Delta\Delta A$ exact sign reversal between enantiomers therefore is an experimental prediction of the rotational strength modulation mechanism.